\documentclass[pdftex,twocolumn,epjc3,final]{svjour3}

\pdfoutput=1 

\usepackage{dblfloatfix}
\usepackage{xcolor}
\usepackage{fixltx2e}

\def\Nch{\ensuremath{N_{\mathrm{ch}}}}
\def\pTzero{\ensuremath{p_{\mathrm{T,0}}}}
\def\vv{\ensuremath{v_2}}
\def\vn{\ensuremath{v_n}}

\usepackage{preprintcover}  

\PreprintCoverPaperTitle{Measurement of the centrality and pseudorapidity dependence of the integrated elliptic flow in lead-lead collisions at $\sqrt{s_{\mathrm{NN}}}=2.76$~\TeV\  with the ATLAS detector}  

\PreprintIdNumber{CERN-PH-EP-2014-065} 

\PreprintCoverAbstract{  The integrated elliptic flow of charged particles produced in Pb+Pb collisions at
  $\sqrt{s_{\mathrm{NN}}}=2.76$~\TeV\
  has been measured with the ATLAS
  detector  using data collected at the Large Hadron
  Collider. The anisotropy parameter, $\vv$, was measured in the pseudorapidity range $|\eta|\leq$ 2.5 with the event-plane method.  
  In order to include tracks with very low transverse momentum \pT,
  thus reducing the uncertainty in $\vv$ integrated over \pT, a $1\,\mu \mathrm{b}^{-1}$ 
  data sample recorded without a magnetic field in the tracking detectors is used. 
  The centrality dependence of the integrated $\vv$ is compared to other
  measurements obtained with higher \pT\ thresholds. A weak
  pseudorapidity dependence of the integrated elliptic flow is
  observed for central collisions, and a small decrease when moving
  away from mid-rapidity is observed only in peripheral collisions. 
  The integrated $\vv$ transformed to the rest frame of one of the
  colliding nuclei is compared to the lower-energy RHIC data. }  

\PreprintJournalName{EPJ C}  

\usepackage{graphicx}
\usepackage{atlasphysics}

\usepackage[T1]{fontenc}
\usepackage{graphicx}
\usepackage{mathptmx}      
\usepackage{lineno}
\usepackage{flushend}
\usepackage[numbers,sort&compress]{natbib}
\usepackage{placeins}
\usepackage[hyperindex,breaklinks]{hyperref} 

\journalname{Eur. Phys. J. C}
\usepackage[normalem]{ulem}
\usepackage{color}

\begin{document}

\renewcommand\linenumberfont{\normalfont\tiny}

\setlength\linenumbersep{1mm}

\title{Measurement of the centrality and pseudorapidity dependence of the integrated elliptic flow in lead-lead collisions at $\sqrt{s_{\mathrm{NN}}}=2.76$~\TeV\  with the ATLAS detector}

\author{The ATLAS Collaboration}
\institute{The ATLAS Collaboration \at CERN, 1211 Geneva 23, Switzerland}

\date{Received: date / Accepted: date}
\titlerunning{Integrated elliptic flow in lead-lead collisions  at
  $\sqrt{s_{\mathrm{NN}}}=2.76$~\TeV }
\maketitle 
\xspaceskip=0.4em
\begin{abstract} 
 The integrated elliptic flow of charged particles produced in Pb+Pb collisions at
  $\sqrt{s_{\mathrm{NN}}}=2.76$~\TeV\ has been measured with the ATLAS
  detector using data collected at the Large Hadron Collider. 
  The anisotropy parameter, $\vv$, was measured in the pseudorapidity range $|\eta|\leq$ 2.5 with the event-plane method.  
  In order to include tracks with very low transverse momentum \pT,
  thus reducing the uncertainty in $\vv$ integrated over \pT, a $1\,\mu \mathrm{b}^{-1}$ 
  data sample recorded without a magnetic field in the tracking detectors is used. 
  The centrality dependence of the integrated $\vv$ is compared to other
  measurements obtained with higher \pT\ thresholds. 
  A weak pseudorapidity dependence of the integrated elliptic flow is
  observed for central collisions, and a small decrease when moving
  away from mid-rapidity is observed only in peripheral collisions. 
  The integrated $\vv$ transformed to the rest frame of one of the
colliding nuclei is compared to the lower-energy RHIC data.  
\end{abstract}

\newlength{\colfig}
\setlength{\colfig}{0.95\columnwidth}

\section{Introduction}
\label{sec:introduction}
The anisotropy in the azimuthal angle distribution of particles
produced in heavy-ion collisions has been studied extensively due to
its sensitivity to the properties of the produced hadronic medium~\cite{Romatschke:2009im, Hirano:2005xf}. 
The final-state anisotropy arises from the initial spatial asymmetry
of the overlap zone in the collision of two nuclei with non-zero
impact parameter. The initial spatial asymmetry induces asymmetric
pressure gradients that are stronger in the direction of the reaction
plane and, due to the collective expansion, lead to an azimuthally
asymmetric distribution of the ejected particles.  
The final-state anisotropy is customarily characterized by the
coefficients $\vn$ of the Fourier decomposition of the azimuthal angle
distribution of the emitted particles~\cite{Poskanzer:1998yz}. 
The second Fourier coefficient $\vv$ is related to the elliptical
shape of the overlap region in non-central heavy-ion collisions,
 and the higher flow harmonics reflect fluctuations in the initial collision geometry~\cite{Aad:2013xma}.  
The first observation of elliptic flow, quantified by measurements
of \vv, at RHIC~\cite{Arsene:2004fa,Back:2004je, Adams:2005dq,Adcox:2004mh} were
found to be well described by predictions based on relativistic
hydrodynamics~\cite{Kestin:2008bh,Chojnacki:2007rq,Romatschke:2007mq},
providing compelling evidence that the created matter is strongly
coupled and behaves like an almost perfect, non-viscous, fluid.
Later studies show small deviations from ideal hydrodynamics,
described in terms of the ratio of shear viscosity to entropy density~\cite{Alver:2010dn, Heinz:2009cv, Song:2009rh, Qiu:2011hf}. 

First results from
Pb+Pb collisions at $\sqrt{s_{\mathrm{NN}}}=2.76$\,\TeV\
\citep{Aamodt:2010pa, ALICE:2011ab,ATLAS:2011ah, ATLAS:2012at,Chatrchyan:2012ta,Chatrchyan:2012xq} from the Large Hadron
Collider (LHC) showed no change in the transverse momentum, \pT,  dependence
of elliptic flow from that measured at the highest RHIC energy, while
the elliptic flow integrated over
\pT\ \cite{Aamodt:2010pa,Chatrchyan:2012ta} was found to increase by
about 30\% from the RHIC energy of $\sqrt{s_{\mathrm{NN}}}=200$\,\GeV\ 
to $\sqrt{s_{\mathrm{NN}}}=2.76$\,\TeV\ at the LHC.  
This increase in the integrated elliptic flow with energy is therefore
driven mostly by the increase in the mean \pT\ of the produced particles.
The dependence of elliptic flow on the geometry of the
collision (the collision centrality) is of particular importance, since
the flow is thought to depend strongly on the initial spatial
anisotropy. Hydrodynamical models are used to quantitatively relate
the initial geometry to the experimentally measured distributions.
Furthermore, recent hydrodynamical
calculations \cite{Schenke:2010rr,Bozek:2011ua} also include a
longitudinal dependence in the source shape, which can be 
deduced from
flow measurements over a wide pseudorapidity range.

This article presents 
measurements of the centrality and
pseudorapidity dependence of the elliptic flow integrated over the \pT\ of
charged particles produced in Pb+Pb collisions at
$\sqrt{s_{\mathrm{NN}}}=2.76$\,\TeV\ recorded in 2010 by the ATLAS
detector.

In order to reduce the uncertainty in the $\pT$-integrated coefficient
$\vv$ by including tracks with \pT\ lower than in the measurements 
reported by the ALICE~\cite{Aamodt:2010pa} and
CMS~\cite{Chatrchyan:2012ta} 
experiments, a special track reconstruction procedure was applied to
``field-off'' data taken without the solenoid's magnetic field in the
tracking detectors.
Other track reconstruction methods, applicable at higher \pT, were
exploited in cross-checks using ``field-on'' data taken with the
solenoid's magnetic field.

\section{The ATLAS detector}
\label{sec:detector}
The ATLAS detector is a multi-purpose particle physics apparatus
and is described in detail elsewhere~\cite{Aad:2008zzm}. 
This analysis uses the three-level trigger system to select
Pb+Pb collision events, the forward calorimeters (FCal) to measure
the collision centrality, and the inner detector (ID) to measure 
charged-particle tracks. 
The ID tracking system comprises silicon pixel and microstrip
detectors and a transition radiation tracker. It provides complete azimuthal
coverage and spans the pseudorapidity region $|\eta| < 2.5$.\footnote{ATLAS uses a right-handed coordinate system with its origin
  at the nominal interaction point (IP) in the centre of the detector
  and the $z$-axis along the beam pipe. The $x$-axis points from the
  IP to the centre of the LHC ring, and the $y$-axis points
  upward. Cylindrical coordinates $(r,\phi)$ are used in the
  transverse plane, $\phi$ being the azimuthal angle around the beam
  pipe. The pseudorapidity is defined in terms of the polar angle
  $\theta$ as $\eta=-\ln\tan(\theta/2)$.} 
The pixel detector consists of a three-layer barrel section 
and three discs in each of the forward regions.  
The semiconductor tracker has four double layers of microstrip sensors in 
its barrel section and nine discs in each of the forward regions. 
The ID is surrounded by a thin superconducting solenoid, which produces a
2\,T axial magnetic field for the field-on data.
The FCal measures both electromagnetic and hadronic energy, using 
copper--tungsten/liquid-argon technology, and provides complete
azimuthal coverage for $3.2<|\eta|<4.9$. 
The hardware-based Level-1 trigger selected minimum-bias Pb+Pb collisions
by requiring either a coincidence of signals recorded in the zero-degree
calorimeters (ZDC) located symmetrically at $z=\pm140$\,m 
($|\eta|>8.3)$ or a signal in at least one side of the minimum-bias trigger scintillators
(MBTS) at $z=\pm3.6$\,m ($2.1<|\eta|<3.9$).
To suppress beam backgrounds, the Level-2 trigger demanded MBTS
signals from opposite sides of the interaction region and imposed a
timing requirement on them.   

With these trigger conditions, 
ATLAS recorded a sample of Pb+Pb collisions corresponding to an
integrated luminosity of approximately $1\,\mu \mathrm{b}^{-1}$ taken
with the field provided by the solenoid turned off.  
In addition, approximately $0.5\,\mu \mathrm{b}^{-1}$ of field-on
data was used in studies of track reconstruction performance.

\section{Event selection and centrality definition}
\label{sec:samples}
The offline event selection required each event to have a vertex
formed by at least three charged-particle tracks reconstructed
in the ID.
The data were recorded at low instantaneous luminosity where the
probability of multiple collisions per bunch crossing (pile-up) was negligible.  
The track reconstruction algorithms therefore allowed only one
collision vertex (called the primary vertex) in each event, thereby
reducing the processing time while maintaining efficiency.
The time difference between the MBTS signals from the opposite sides
of the interaction region was required to be less than 3\,ns, and 
a coincidence of ZDC signals was also required. 
These additional selection criteria efficiently remove beam-gas and 
photo-nuclear interactions. 
As shown in previous studies~\cite{ATLAS:2011ah}, the applied trigger
and offline requirements provide a minimum-bias event sample in which 
the fraction of inelastic Pb+Pb collisions is $98\pm 2$\%.  

Events satisfying the above criteria were also required to have a
primary vertex within 50\,mm (100\,mm) in the z-direction of the nominal centre of the
ATLAS detector for the field-off (field-on) data subsample. 
After requiring all relevant subdetectors to be performing normally,
the subsamples used in the analysis of the field-off and field-on data
contained approximately 1.6 million and 3 million minimum-bias
events, respectively.

Monte Carlo (MC) event samples were used to determine the tracking
efficiencies and the rates of fake tracks.
The HIJING event generator~\cite{Gyulassy:1994ew} was used to produce 
minimum-bias Pb+Pb collisions. 
Events were generated with the default parameters except for jet
quenching, which was turned off. 
The effect of elliptic flow was implemented after event generation.
The azimuthal angles of final-state particles were redistributed at
generator level to produce an elliptic flow consistent with previous
ATLAS measurements~\cite{ATLAS:2011ah,ATLAS:2012at}. 
The simulation of the ATLAS detector's response~\cite{ATLAS:2010ah}
to the generated events is based on the GEANT4
package~\cite{Agostinelli:2002hh} and included a detailed description
of the detector geometry and material in the 2010 Pb+Pb run.
Two samples of 0.5 million MC events were simulated, one with the
solenoid field switched off and the other with it switched on. 
Additional MC samples consisting of 50,000 events simulated with
10--20\% extra detector material were used to study systematic uncertainties.
The generated charged particles
were reweighted with \pT- and centrality-dependent functions so that
the \pT\ spectra in the MC samples matched the experimental ones~\cite{ATLAS:2011ag}.

The centrality of the Pb+Pb collisions was characterized by the summed
transverse energy,  {\mbox{$\Sigma E_{\mathrm{T}}^{\mathrm{FCal}}$}}, measured
in the FCal~\cite{ATLAS:2011ah}. 
The {\mbox{$\Sigma E_{\mathrm{T}}^{\mathrm{FCal}}$}} distribution was
divided into ten centrality bins, each representing 10\% of the full
distribution after accounting for 2\% inefficiency in recording the
most peripheral collisions (the 0--10\% centrality interval 
corresponds to the most central 10\% of collisions: those with the
largest {\mbox{$\Sigma E_{\mathrm{T}}^{\mathrm{FCal}}$}}). 
A small change in the overall recording efficiency leads to large
fluctuations in the population of the most peripheral collisions.
To avoid resulting large systematic uncertainties, the 20\% of events
with the smallest {\mbox{$\Sigma E_{\mathrm{T}}^{\mathrm{FCal}}$}}
were not included in the analysis.

\section{Elliptic-flow measurement}
\label{sec:flow}
The final-state azimuthal anisotropy is quantified by the coefficients
in the Fourier expansion of the $\phi$ distribution of charged particles
\cite{Poskanzer:1998yz},
\begin{equation}
  \centering
  {\mathrm{d}}N/{\mathrm{d}}\phi \propto 1+2\sum_{n=1}^{\infty}\vn\cos{(n[\phi-\Psi_n])},
  \label{eg:flow}
\end{equation}
where $\vn$ and $\Psi_n$ are the magnitude and the azimuthal direction
(called the event-plane angle) of the $n$-th flow harmonic, respectively. 

The second flow harmonic, $\vv$, for a given collision centrality is a function of \pT\ and $\eta$,
and is determined by
\begin{equation}
\vv(\eta,\pT) =\frac{\langle\cos{(2[\phi-\Psi_2])}\rangle}{\sqrt{\langle \cos{(2[\Psi_2^N-\Psi_2^P])}\rangle}},
\label{eq:v2}
\end{equation}
where the numerator denotes the average over charged-particle tracks
in a given $\eta$ and \pT\ range, and the denominator, averaged over
events, is a correction accounting for the finite experimental
resolution in the determination of the event-plane angle $\Psi_2$. 
This resolution correction was obtained using the sub-event
technique~\cite{Poskanzer:1998yz}  as described in
Refs~\cite{ATLAS:2011ah,ATLAS:2012at}. The two ``sub-events'' defined for each event
cover two $\eta$ ranges of the same width in the positive and
negative $\eta$ hemispheres ($3.2<|\eta|<4.8$) of the FCal
detector. The sub-event-plane angles are determined by
\begin{equation}
\label{eq:eptracks}
\Psi_2^{\mathrm{N(P)}} = \frac{1}{2}\tan^{-1}\left( \frac{\sum\limits_i E^{\mathrm{tower}}_{\mathrm{T}i} w_i\sin(2\phi_i)}
{\sum\limits_i E^{\mathrm{tower}}_{\mathrm{T}i} w_i\cos(2\phi_i)}\right),
\end{equation}
\noindent
where the sums run over transverse energies,
$E_\mathrm{T}^{\mathrm{tower}}$, as measured in calorimeter towers
located at negative (N) and positive (P) $\eta$ in the first sampling
layer of the FCal. 
The FCal towers consist of
cells grouped into projective regions in 
$\Delta \eta \times \Delta \phi$ of $0.1 \times 0.1$.  
The weights, $w_i = w_i(\phi_i,\eta_i)$, were determined in narrow $\Delta\eta$ slices
($\Delta \eta = 0.1$) over the full $\eta$ range in order to remove
structures in the uncorrected $\phi$ distributions of
$E_\mathrm{T}^{\mathrm{tower}}$ in every $\Delta\eta$ slice and to
ensure a uniform $\Psi_2$ distribution. 

In the sub-event approach, potential non-flow correlations are
minimized by using the reaction plane estimated from the 
$\eta$ side opposite to the tracks used for the $\vv$ measurement;
this provides a separation of $\Delta\eta > 3.2$. 
This method was previously applied~\cite{ATLAS:2011ah} to
measure $\vv$ as a function of \pT\ using charged-particle tracks
reconstructed in the ID tracking system with a minimum \pT\ of
0.5\,\GeV. 

In order to perform the integration over \pT, the differential $\vv$
measurements are weighted by the number of charged-particle tracks $N_{i,k}^{\mathrm{corr}}$,
\begin{equation}
\label{eq:v2integrated}
 \vv = \sum_i \sum_k\vv(\eta_i,p_{\mathrm{T},k}) N_{i,k}^{\mathrm{corr}}/\sum_i \sum_k{N_{i,k}^{\mathrm{corr}}},
 \end{equation}
 \noindent
and summed over bins in $\eta$ (denoted by the index $i$)
 and \pT\ (index $k$). The number of charged-particle tracks is
 calculated as
$ N_{i,k}^{\mathrm{corr}} =N_{i,k}[1-f(i,k)]/\epsilon(i,k), $
where the $N_{i,k}$ is the observed number of tracks in a given $\eta$ and \pT\ bin,
$\epsilon(i,k)$ is the track reconstruction efficiency and $f(i,k)$ is
the estimated rate of fake tracks.
In the following sections, the lower limit in the integration of $\vv$ over \pT\ is denoted by \pTzero.

\section{Track reconstruction}
\label{sec:reconstruction}
The ID was used to reconstruct charged-particle trajectories.  
Three track reconstruction methods were applied in 
order to exploit a large range in particle \pT:
\begin{figure}[t]
  \begin{center}
    \includegraphics[width=\colfig]{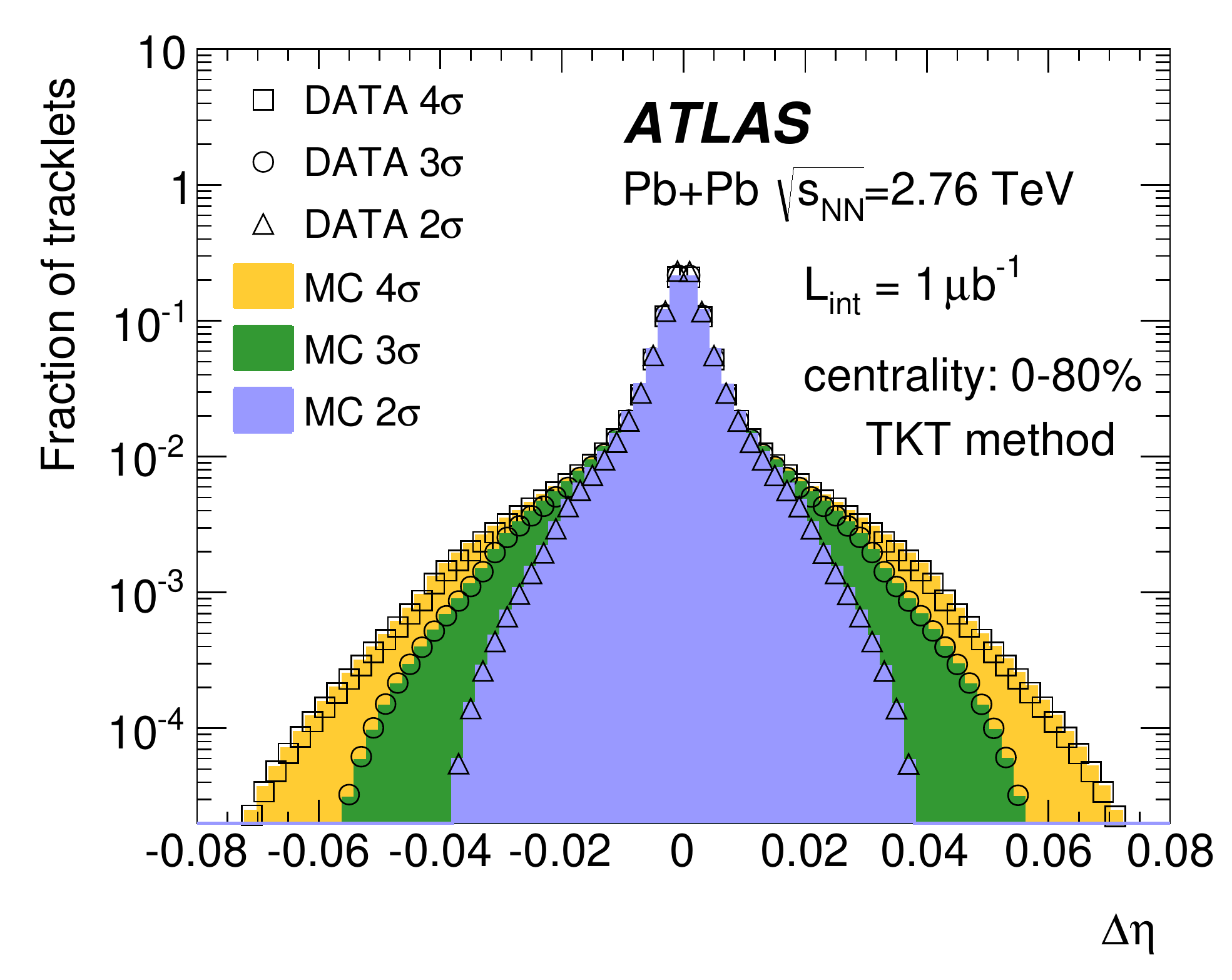}\\
    \includegraphics[width=\colfig]{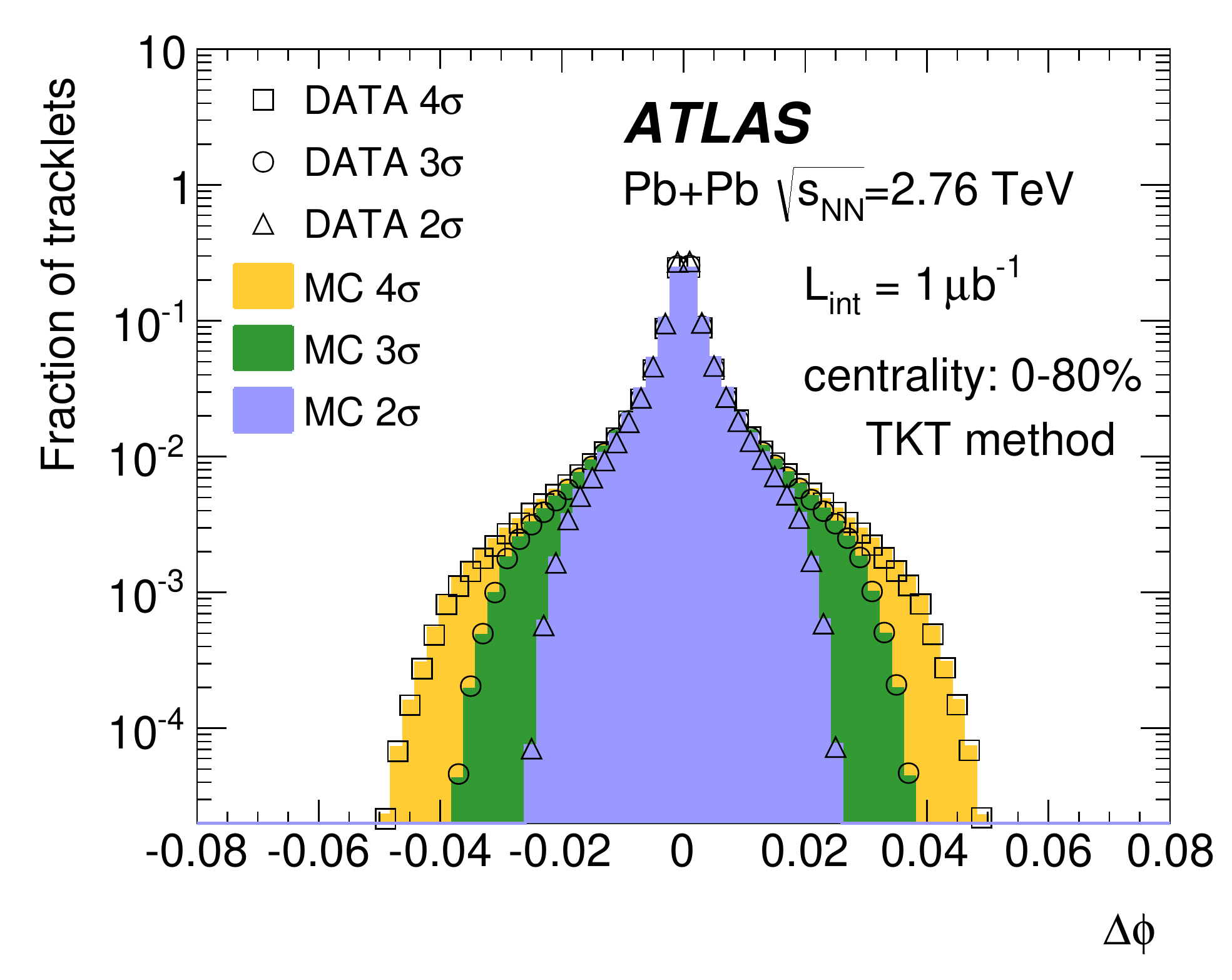}\\
    \caption[tracklet performance]{Comparison of the tracklets'
      $\Delta\eta$ (top) and $\Delta\phi$ (bottom) distributions in data
      (open symbols) and MC simulation (filled histograms) for
      tracklets measured within the pseudorapidity range $|\eta|<2$,
      for events in the 0--80\% centrality interval and $\Delta R < 4 \sigma, 3
      \sigma$ and $2\sigma$ (see Sect.~\ref{sec:reconstruction} for details) as described in the legend.  }
    \label{fig:trackletPerfDataMCComp}	
  \end{center}
\end{figure}
\begin{itemize}
  \item the tracklet (TKT) method used for the field-off data in order to reach
    charged-particle \pT\ below 0.1\,\GeV~\cite{ATLAS:2011ag},

\item  the pixel track (PXT) method used to reconstruct tracks with $\pT \geq 0.1$\,\GeV\ using only the pixel
  detector in the field-on data sample,

\item   the ID track (IDT) method for the field-on data sample, the default
  ATLAS reconstruction method, for which all ID sub-detectors are used
  and the track \pT\ is limited to $\pT \geq 0.5$\,\GeV~\cite{Aad:2010rd}.

\end{itemize}
\begin{figure}[tb]
\begin{center}
\vspace*{2.8mm}
\includegraphics[width=\colfig]{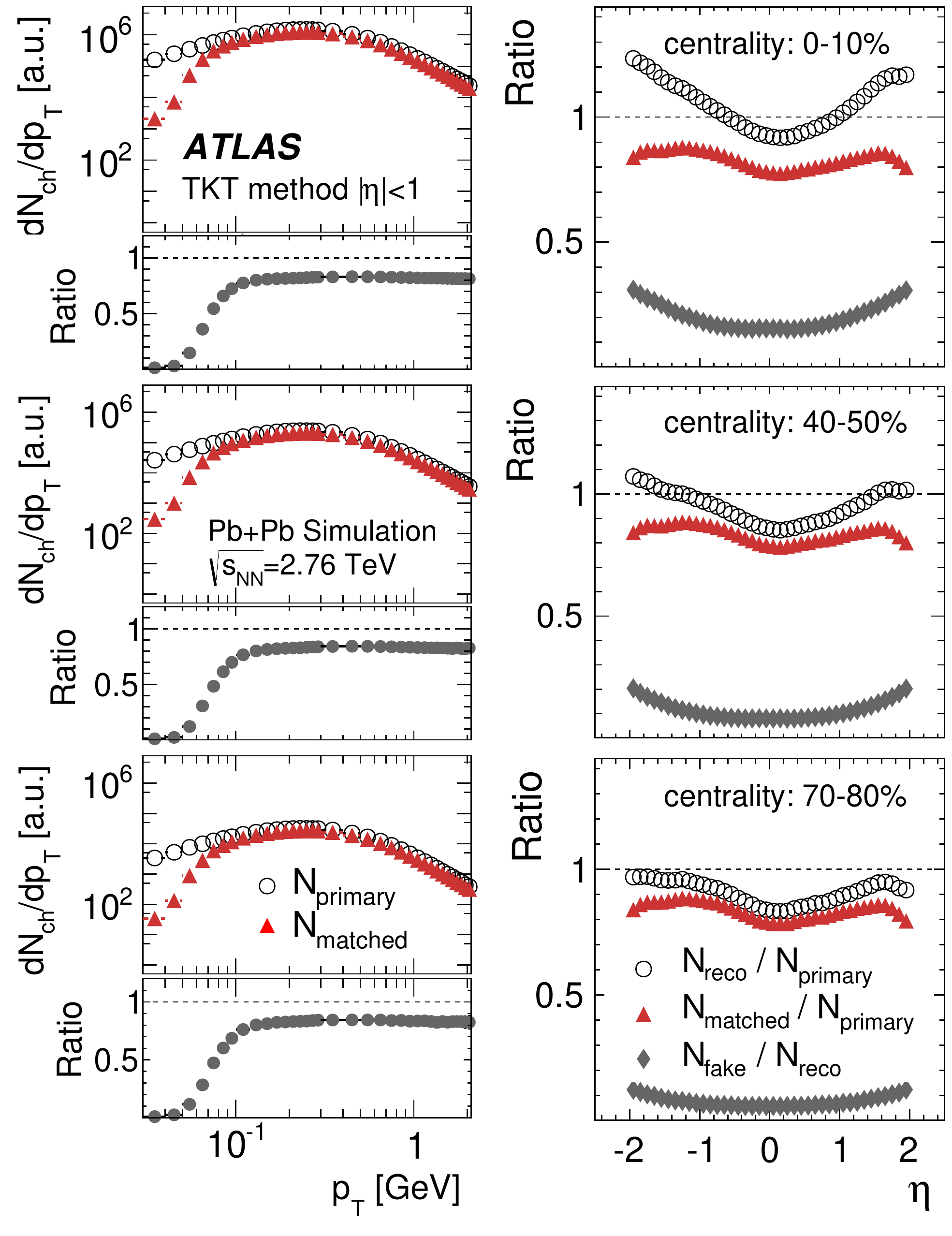}
\caption[tracklet performance]{\label{fig:trackletPerf} Monte Carlo evaluation
  of the tracklet reconstruction performance in representative
  centrality bins 0--10\%, 40--50\% and 70--80\%. 
  Left: generator-level transverse momentum
  distributions of primary charged particles, $N_{\mathrm{primary}}$ (open circles),
  compared to the \pT\ spectra of charged particles matched to the
  reconstructed tracklets, $N_{\mathrm{matched}}$ (red triangles). Bottom panels show
  the ratios of the two distributions. Right: pseudorapidity, $\eta$, 
  dependence of the ratio of all reconstructed tracklets,
  $N_{\mathrm{reco}}$ (open circles), and $N_{\mathrm{matched}}$ (red triangles)
  to all primary charged particles. The ratio of
  fake tracklets, $N_{\mathrm{fake}}$ (grey diamonds), to all reconstructed
  tracklets is also shown.
}
\end{center}
\end{figure}
In the TKT method for field-off data, tracks are formed from the
positions of hit clusters in the inner two layers of the pixel detector
and the position of the primary vertex reconstructed using ID tracks.
In the first step,
the $\eta_0$ and $\phi_0$ coordinates are defined using the event's vertex
position and the hit recorded in the first pixel layer. Then a
search for a hit in the second pixel layer (with $\eta_1$ and
$\phi_1$ coordinates defined with respect to the vertex position) is
performed and its consistency with a straight-track hypothesis is
checked. Candidate tracklets are required to satisfy the condition
\begin{equation}
\label{eq:tracklet}
\Delta R = \frac{1}{\sqrt{2}} \sqrt{\left( \frac{\Delta\eta}{\sigma_{\eta}(\eta_0)} \right)^{2} + \left( \frac{\Delta\phi}{\sigma_{\phi}(\eta_0))} \right)^{2}} < N_{\sigma},
\end{equation}
where $\Delta\eta=\eta_{1}-\eta_{0}$ and $\Delta \phi=\phi_1-\phi_0$,
and $\sigma_{\eta}(\eta_0)$ and $\sigma_{\phi}(\eta_0)$ are
pseudorapidity-dependent widths of the $\Delta\eta$ and $\Delta\phi$
distributions, respectively. In this analysis, $N_{\sigma}=3$ was used
as the default condition.  
Clusters located close to each other in the second pixel layer are most
likely to originate from the same particle. 
Therefore, if more than one cluster located in the second pixel layer
fulfils the selection criteria, the resulting tracklets are merged
into a single tracklet. 
The $\Delta \eta$ and $\Delta \phi$ distributions in data and MC
simulation are compared in Fig.~\ref{fig:trackletPerfDataMCComp}.
The data and MC distributions agree well.  Candidates fulfilling the criterion
in Eq.~(\ref{eq:tracklet}) were accepted for further analysis with
$\eta=\eta_0$ and $\phi=\phi_0$.

This method does not provide information about the track's \pT;
nevertheless, its performance can be checked as a function of
generator-level particle \pT\ by applying the same reconstruction 
procedure to the MC simulation and using the \pT\ of the generated
particle corresponding to the reconstructed tracklet whenever applicable.
Figure~\ref{fig:trackletPerf} compares the \pT\ spectra of stable
charged particles at the MC-generator level, $N_{\mathrm{primary}}$,
to the spectra of reconstructed tracklets matched to charged
particles, $N_{\mathrm{matched}}$, for three representative 
centrality bins and for $|\eta| < 1$. 
A particle was considered to be primary if it originated directly from
the collision or resulted from the decay of a particle with $c\tau<1$\,mm.  
The matching criterion required that the two hits forming the tracklet
be identical to the hits associated with a charged particle. 
The distributions show that the TKT method is able to reconstruct
particles with transverse momenta $\sim 0.07$\,\GeV\ with 50\%
efficiency, and that a plateau at about 80\% is reached for 
$\pT > 0.1$\,\GeV\ in all centrality bins.
For low \pT, the efficiency decreases sharply, but the particle
density is small in this region, as is $\vv$; thus the contribution from
this region to the integrated elliptic flow is expected to be small.  
Figure~\ref{fig:trackletPerf} also shows the reconstruction efficiency,
$N_{\mathrm{matched}}/N_{\mathrm{primary}}$, as a function of $\eta$. 
Here, $N_{\mathrm{primary}}$ denotes all primary charged particles
with $\pT \ge 0.07$\,\GeV, which defines the low-\pT\ limit for
integrating $\vv$ over \pT . 
The efficiency is found to be  $\sim 80$\% and depends weakly on
$\eta$. 
The rate of fake tracklets, $N_{\mathrm{fake}}$, measured as the ratio
of the number of tracklets not matched to charged particles to the
total number of reconstructed tracklets,
$N_{\mathrm{fake}}/N_{\mathrm{reco}}$, increases with centrality and
$|\eta|$, reaching about 35\% for the most central collisions at
$|\eta|=2$.

\begin{figure}[t]
  \centering
  \includegraphics[width=\colfig]{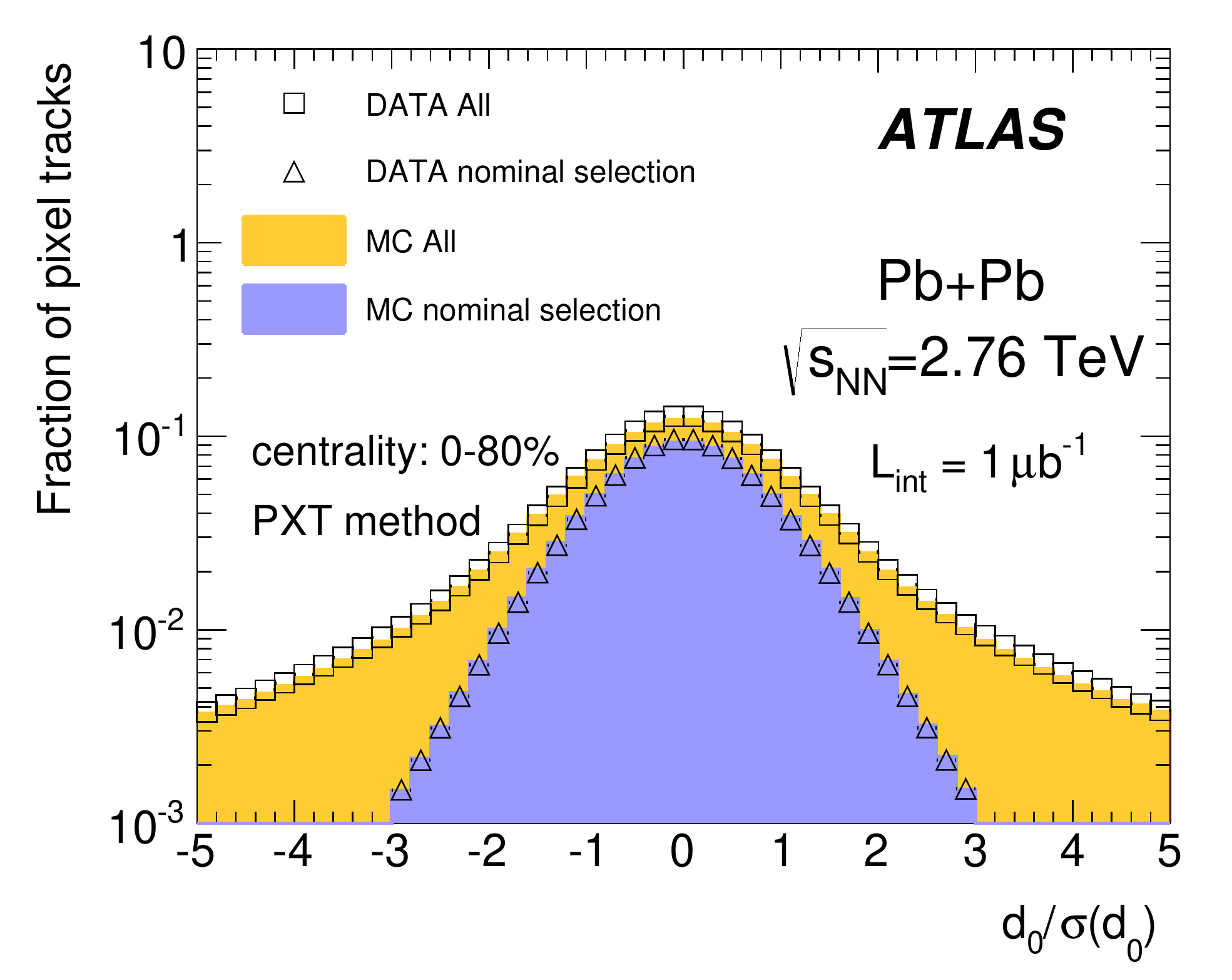}\\
  \includegraphics[width=\colfig]{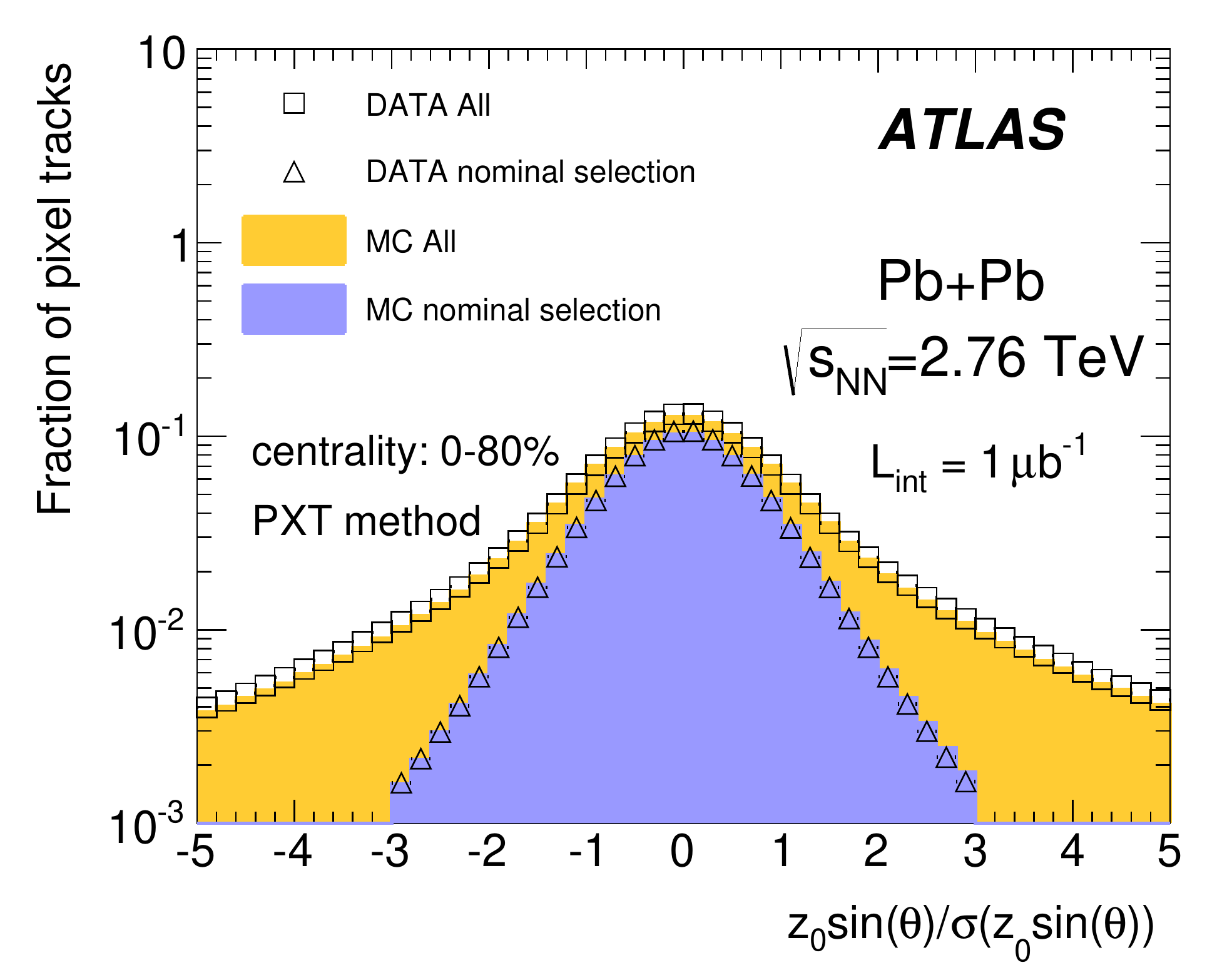}\\
  \caption[pixel performance]{\label{fig:significances} Comparison of
    distributions of the transverse (top), and longitudinal (bottom) impact parameter
    significances in data and MC simulation for all reconstructed tracks
    and for the selected tracks (see text for details). }
\end{figure}
For field-on data, the PXT method allows the transverse momentum range
$\pT > 0.1$\,\GeV\ to be examined. 
Tracks were reconstructed within the full acceptance of the pixel
detector ($|\eta|<2.5$). 
To improve the track reconstruction's performance in the heavy-ion
collision environment, the track-quality requirements were made more
stringent than those for proton--proton collisions~\cite{Aad:2010ac}.
Pixel tracks were required to have no missing hits in the pixel
layers, and the transverse and longitudinal impact parameters,  $d_0$
and  $z_0$, with respect to the vertex
were required to have
$|d_0|$ and $|z_0\sin(\theta)|$ less than 1\,mm and significances
$|d_0/\sigma_{d_0}|$ and $|z_0 \sin{\theta}/\sigma_{z_0  \sin{\theta}}|$ less than 3.0. 
Figure~\ref{fig:significances} shows good agreement between data and
MC simulation in the distributions of $|d_0/\sigma_{d_0}|$ and 
$|z_0\sin{\theta}/\sigma_{z_0 \sin{\theta}}|$.

\begin{figure}[!thb]
\begin{center}
\vspace*{2.8mm}
\includegraphics[width=\colfig]{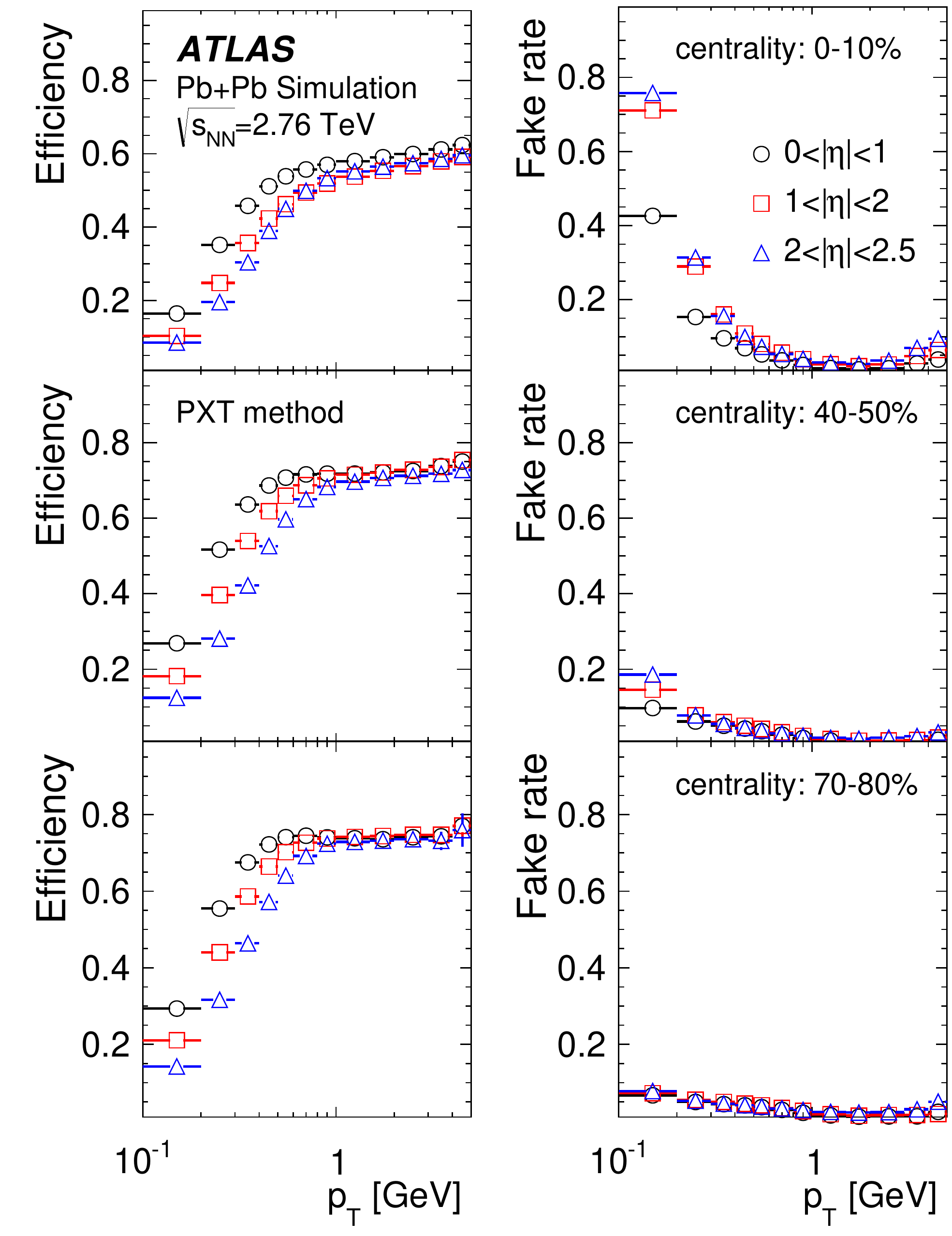}
\caption[pixel performance]{\label{fig:pixelPerf} 
  The transverse momentum, \pT, dependence of the pixel track reconstruction efficiency
  (left) and the fake rate (right) for three pseudorapidity ranges and 
  three centrality intervals as indicated in the legend.}
\end{center}
\end{figure}

The pixel track method's reconstruction efficiency was evaluated in MC
simulation by matching reconstructed tracks to the generated charged
particles. 
A track is matched to a generated charged particle if it is reconstructed from at least 69\% of the pixel hits originating from the latter.
Figure~\ref{fig:pixelPerf} illustrates the dependence of the pixel
track reconstruction efficiency on \pT\ in three pseudorapidity ranges
and for three selected centrality bins. 
The efficiency decreases slightly from peripheral to central collisions
and also deteriorates when moving away from mid-rapidity. A weak \pT\
dependence is observed above $\pT > 0.5$\,\GeV\ for all collision
centralities. At lower \pT, the efficiency decreases with decreasing
\pT\ and to about 20\% at the lowest accessible \pT.

The fraction of fake tracks, defined as the ratio of reconstructed
tracks not matched to generated charged particles to all reconstructed
pixel tracks, was evaluated using MC simulation.  
Figure~\ref{fig:pixelPerf} shows the fake-rate dependence on \pT\ in
three pseudorapidity ranges and for three centrality bins.
The fake rate is below 10\% for \pT\ above 0.4\,\GeV\ and depends very
weakly on \pT\ and $\eta$ for peripheral collisions. 
In more central collisions, the fake rate increases at low
\pT\ and shows a similar increase with increasing $|\eta|$.

\begin{figure}[t]
   \begin{center}    
     \vspace*{2.8mm} 
     \includegraphics[width=\colfig]{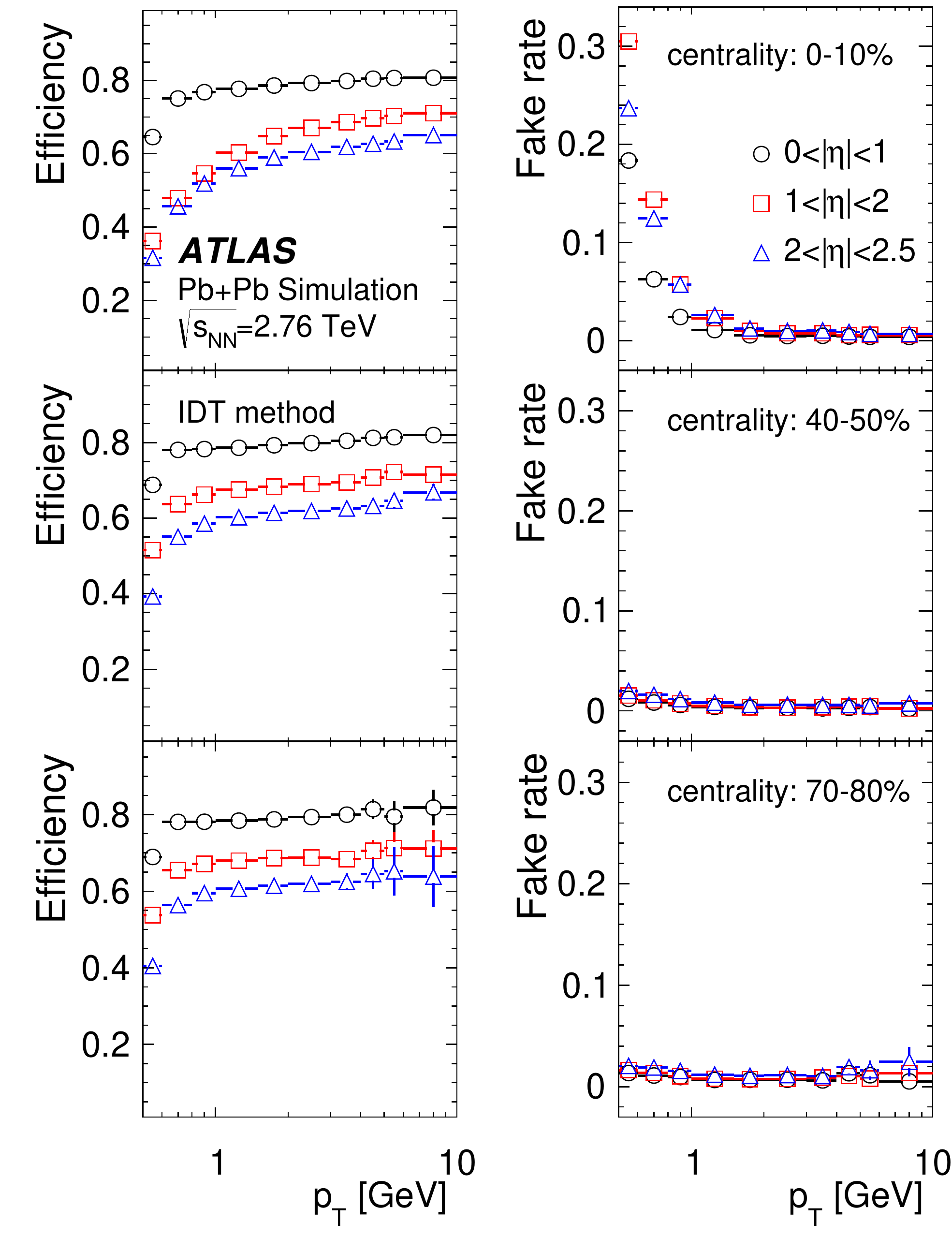}
     \caption[pixel performance]{\label{fig:trackPerf} The transverse
       momentum, \pT, dependence of the ID track reconstruction efficiency
       (left) and the fake rate (right) for three
       pseudorapidity ranges and three centrality intervals as indicated in the legend. }
   \end{center}
 \end{figure}

 The performance of the PXT reconstruction method can be compared
 with that of the IDT method. The track reconstruction efficiency and
 rate of fake tracks from the IDT method are shown in
 Fig.~\ref{fig:trackPerf} (for reconstruction details see
 Ref.~\cite{ATLAS:2011ah}). 
 The minimum \pT\ reached is 0.5\,\GeV.  
A comparison of Figs~\ref{fig:pixelPerf}~and~\ref{fig:trackPerf} shows that
the extension towards lower \pT\ values for the PTX method is achieved
at the expense of much larger fake rates than observed for the IDT
method, whereas the reconstruction efficiencies are similar.  
 The two methods have different \pT\ resolutions: it is very good for  
 ID tracks, the root mean square of
 $(\pT^{\mathrm{reco}}/\pT^{\mathrm{true}}-1)$ being, in $|\eta|<1$, about 4\% and
 independent of the track \pT\ in the used range, whereas for pixel
 tracks it is about 10\% at the lowest $\pT$ and increases to about
 15\% at 5\,\GeV.

 \begin{figure}[t]
   \begin{center}     
     \vspace*{2.8mm}
     \includegraphics[width=\colfig]{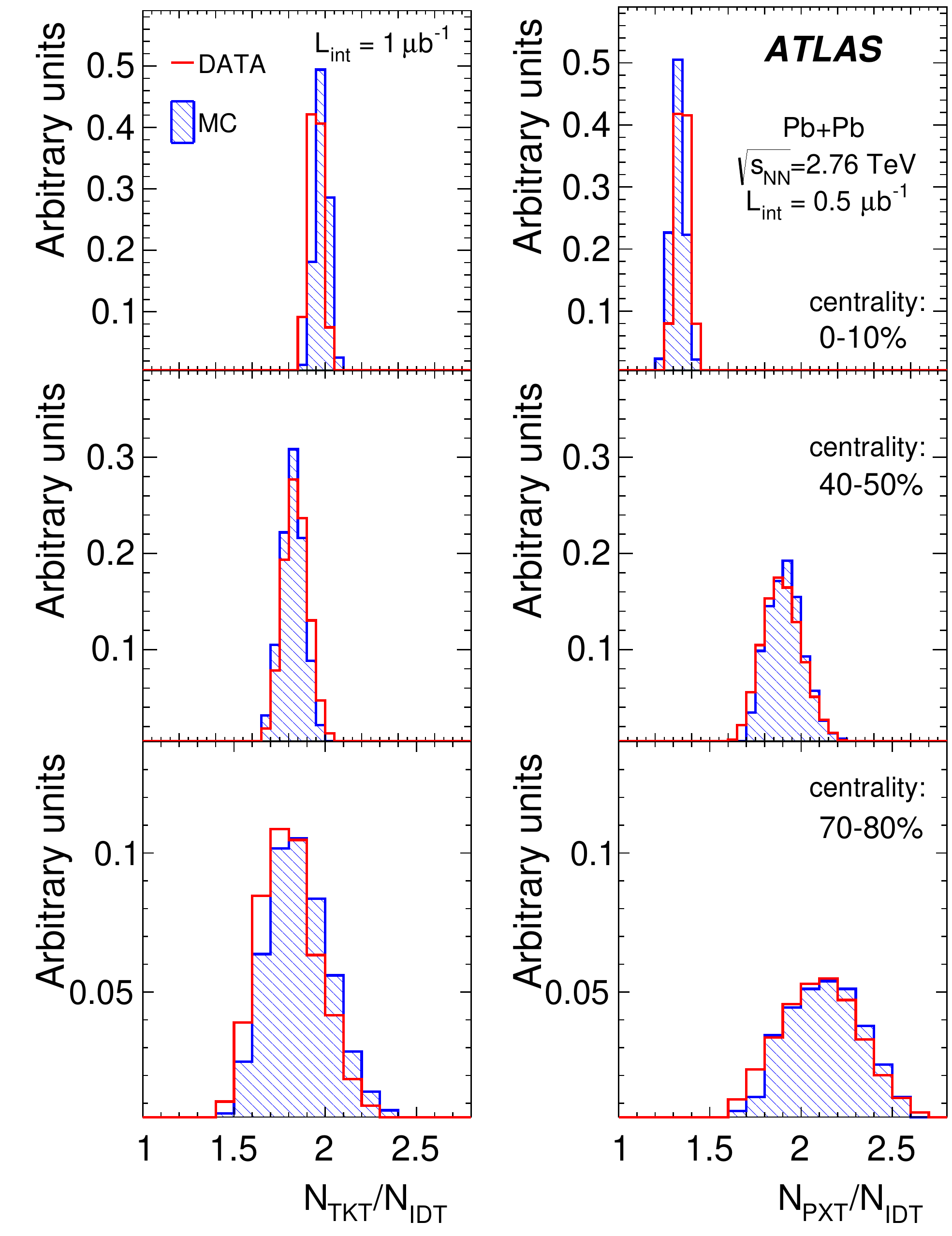}
     \caption[pixel performance]{\label{fig:mult-ratios} Comparison of the
       distribution of multiplicity ratios of number of tracklets,
       $N_\mathrm{TKT}$, (left) and pixel tracks, $N_\mathrm{PXT}$, (right)
       to the number of ID tracks, $N_\mathrm{IDT}$, in data (red) and MC
       simulation (blue) in three centrality bins as indicated on the
       plots. }
   \end{center}
\end{figure}
The performance of the MC simulation in describing the fake rates in
the data was checked by comparing the $\Delta \eta$, $\Delta \phi$,
$d_0/\sigma_{d_0}$ and $z_0 \sin{\theta}/\sigma_{z_0\sin{\theta}}$
distributions, like the ones shown in Figs~\ref{fig:trackletPerfDataMCComp}
and~\ref{fig:significances}. Additionally, the distributions of the
ratios of the number of tracklets and pixel tracks to the number of ID
tracks in data and MC simulation were compared, as shown in
Fig.~\ref{fig:mult-ratios}. It can be concluded that the MC description 
of the TKT and PXT methods' performance is adequate.

The elliptic flow depends on the particle type~\cite{Abelev:2013vea},
as does the reconstruction efficiency.  
Although the track reconstruction efficiency is averaged over all
particle types in this analysis, the reconstruction efficiencies for
simulated pions, kaons and protons are shown as a function of \pT\ 
in the Appendix. This information is important for comparison of
measurements with theory predictions in which the elliptic flow 
depends on the particle type.
\section{Corrections to measured $\vv$}
The event-plane method~\cite{Poskanzer:1998yz} is applied to
measure the differential elliptic flow harmonic $\vv(\eta)$ in small
$\eta$ bins with the TKT method, and $\vv(\eta,\pT)$ in small $\eta$
and \pT\ bins with the PXT and IDT methods. 
The differential $\vv$ measurements are then corrected for
detector-related effects.

The first correction is associated with the variation in tracking
efficiency induced by the flow itself. It is applied only to the PXT
method.  The efficiency of the reconstruction was found to depend on
detector occupancy, which in turn scales with the particle
multiplicity. 
Since the flow phenomenon is a modulation of the multiplicity, it may 
induce a variation of the tracking efficiency in an event.  
Higher occupancy causes lower efficiency, and the number of tracks
observed in the event plane is reduced more strongly than the number
of tracks observed in other directions.  
As a consequence, the observed $\vv$ is smaller.  
In order to correct for this effect, an appropriate weight was applied
to the tracks in the calculation of the numerator of~Eq.~(\ref{eq:v2}).  
This weight, the inverted efficiency parameterized as a function of
detector occupancy in the vicinity of the track, was derived from MC
simulation.  
In the data, the occupancy was determined for each track from the
number of hits near the track in the first layer of the pixel detector.  
The corrected $\vv(\pT)$ was compared to the measurement obtained from
the IDT method in the same data. 
In the MC simulation, the comparison was made to $\vv(\pT)$ determined
using generated particles. 
The relative increases in the value of $\vv(\pT)$  in data and in simulation
were found to be compatible for $\pT > 0.5$\,\GeV, the range in which
the comparison could be performed. 

The occupancy correction results in an increase of about 12\% in
the integrated $\vv$ for the 0--20\% centrality interval while it
amounts to only 1\% for the most peripheral collisions, when using a
lower \pT\ integration limit of $\pTzero = 0.1$\,\GeV. 
For higher values of \pTzero\ the correction gradually becomes smaller. 
For $\pTzero = 0.5$\,\GeV\ it decreases to about 7\% for the most
central collisions.

An additional correction, applied to the differential measurement of
$\vv$, accounts for the difference between $\vv$ measured only with
fake tracks and $\vv$ measured with charged-particle tracks from the
primary vertex. 
The corrected $\vv$ is calculated as
\begin{equation}
\vv = \frac{\vv{}_{,m} - f\vv{}_{,f}}{1-f},
\end{equation}
where $\vv{}_{,m}$ is the elliptic flow measured with all tracks, $\vv{}_{,f}$ is
the flow measured with fake tracks, and $f$ is the fake-track rate. 
This correction was applied to the differential $\vv$ measured with
the TKT, PXT and IDT methods with the corresponding fake rates 
and $\vv{}_{,f}$ values.  
The rate and $\vv{}_{,f}$ of the fake tracks were derived from MC simulation and
then cross-checked in the data with a sample, obtained with inverted
track selection criteria, in which fake tracks dominate.  
Differences between the MC simulation and the data of up to 20\% were 
observed and included in the systematic uncertainties.

The fake tracks reduce the values of $\vv$  integrated over the \pT\
ranges considered in this analysis.   
The correction is a function of the fake-track rate and accordingly
exhibits a dependence on centrality, \pT\ and $\eta$. 
For $|\eta|<1$, the largest correction, about 15\%, was obtained for
the PXT method with $\pTzero = 0.1$\,\GeV. 
For peripheral collisions in the same kinematic range, it decreases
to about 11\%. 
The correction is smaller for higher values of \pTzero. 
It decreases to about 2\% for $\pTzero = 0.5$\,\GeV\ for the 0--10\%
centrality interval and gradually drops to zero for the most
peripheral collisions. 
The fake-track flow correction for the integrated $\vv$ obtained with
the IDT method ($\pTzero = 0.5$\,\GeV) is less than 2\% for the most
central collisions and even smaller for the more peripheral ones. 
For the TKT method, the correction is about 1\% for the most central 
collisions.

The limited \pT\ resolution for tracks reconstructed in the pixel
detector and the rapidly changing ${\mathrm{d}}\Nch/{\mathrm{d}}\pT$
distribution lead to a significant bin-to-bin migration in \pT.  
As a consequence of the variation of $\vv$ with \pT,  $\vv$ measured
in a given \pT\ bin is contaminated by $\vv$ values of particles from
the neighbouring bins.  
In order to compensate for this effect, a correction derived from MC
simulation was applied to the $\vv(\pT)$ values.  
This correction was determined, using pixel tracks matched to
generated particles, by comparing the $\vv(\pT)$ distribution as a
function of reconstructed \pT\ to  $\vv(\pT)$ as a function of
generated \pT. 
In order to validate the correction derived from the MC simulation,
the same procedure was applied in the data and in the simulation 
in the region of $\pT > 0.5 \GeV$, where the ID tracks were used instead of the generated particles. 
The ID tracks were matched by requiring an angular separation $\sqrt{(\Delta\eta)^2 + (\Delta\phi)^2} < 0.02$.  
A comparison between the  corrections obtained in the data and in the MC simulation, as a function of measured \pT,  showed a good agreement.

The correction for \pT-bin migration of the reconstructed tracks was
found to be small compared to the occupancy and fake-track flow
corrections, and to depend only on the value of \pTzero. 
It increases the integrated $\vv$ value by 1\%  (1.5\%) for 
$\pTzero = 0.1$\,\GeV ($\pTzero = 0.5$\,\GeV) independently of
collision centrality.  

\section{Uncertainties in the $\vv$ determination}

The systematic uncertainties include those common to different
tracking methods, as well as method-specific ones. 

The uncertainty which originates from the statistics of the MC samples
is treated as a source of systematic uncertainty.

The $\vv$ values determined for samples enriched in fake tracks  in
data and MC simulation were compared and differences of up to 20\% for
both the PXT and IDT methods were observed.
For the PXT method, this difference resulted in a change of $\vv$,
integrated from $\pTzero = 0.1$\,\GeV, for the most central (0--10\%)
collisions of 3\% at mid-rapidity and of 15\% at $|\eta|\sim2$.
The impact on the integrated $\vv$ decreases with increasing
centrality.  
For higher $\pTzero$ values, the change was found to be negligible.
For the IDT method, the uncertainty on the $\vv$ value of fake tracks
induces a systematic uncertainty in the integrated $\vv$ for central
collisions of less than 4\% at mid-rapidity and of 9\% at
$|\eta|\sim2$; for peripheral collisions the uncertainty is smaller. 

The variation of the fake tracklets' $\vv$, at the level of 10\%,
obtained from the comparison of data and MC simulation, results in an
uncertainty at the level of 2\% in the integrated $\vv$ across the centrality
range 0--40\%.

A comparison of $\vv$ values obtained with the TKT method for a MC
sample with the nominal detector geometry to that with 10\% more
active material and 15--20\% more inactive material shows agreement to
better than 2\%. 
Therefore it was assumed that possible inaccuracies in the
description of the detector material in the GEANT4 simulation have
a negligible effect on the final results. 
The same holds for the measurements with the PXT and IDT methods.

An overall scale uncertainty on $\vv$ originates from the uncertainty
on the fraction of the total inelastic cross section accepted by the
trigger as well as from the event selection criteria, which affects the population of the centrality bins. 
It is negligibly small (below 1\%) for central collisions and
increases to about 6\% for the most peripheral collisions for the TKT
method and to about 5\% for both the PXT and IDT methods.

The influence of the detector nonuniformities on the measured \vv\ 
was checked by comparing the \vv\ values obtained 
for positive and negative $\eta$. 
This led to a typical uncertainty of 1\% except for the most
peripheral collisions where it increased to about 2\%.

Deviations of $\left<\sin{2[\phi-\Psi_2]}\right>$ from zero point to
detector non-uniformities and biases in the event-plane determination.  
The magnitude of the sine term relative to the cosine term is included
in the systematic uncertainty of $\vv$. 
For the TKT method, its contribution to the relative systematic
uncertainty is negligibly small. 
For the PXT and IDT methods, it is small for most centrality bins,
and increases to 2\% only for the most peripheral collisions.

The analysis procedure was checked with MC studies in which the generated
elliptic flow signal was compared to the $\vv$ values obtained with
the same analysis chain as used for the data. 
In this MC closure test, relative differences of up to 2\% in central
collisions and of up to 5\% in peripheral collisions were observed for
the TKT method. 
For the IDT method, the relative difference reaches 2\%; for the PXT
method, it remains within 2\% except for the most peripheral
collisions where it increases to 5\%. 
The relative difference between the expected and measured values is
included in the total systematic uncertainty.

The $\Delta R$ parameter used in the tracklet reconstruction was varied by  $\pm 1
\sigma$ from the nominal value. The resulting variation in the value
of $\vv$ at the level of 1\% is included in the systematic
uncertainty. For the PXT and IDT methods, differences between $\vv$ determined from
tracks of negatively and positively charged particles as well as
between the baseline $\vv$ and that obtained with tighter or looser
tracking requirements (in which the transverse and longitudinal
impact parameter significance criteria are changed by $\pm 1$) also
contribute to the systematic uncertainty at the level of a few percent.

For the PXT method, the corrections due to the limited \pT\ resolution were varied
within their statistical uncertainties and the resulting variation was
found to be at the level of 0.5\%, independently of the centrality.

The $\pT$ spectrum of charged particles in the MC simulation was
reweighted so that the expected detector-level distribution agrees
with that observed in the data.  
This changes the effective fake-track rate and therefore the weights
used in the calculation of $\vv$. 
A variation of these weights by up to 10\% has a negligible effect on the
determination of $\vv$.

\begin{figure*}[!hb]
\begin{center}
\includegraphics[width=0.9\textwidth]{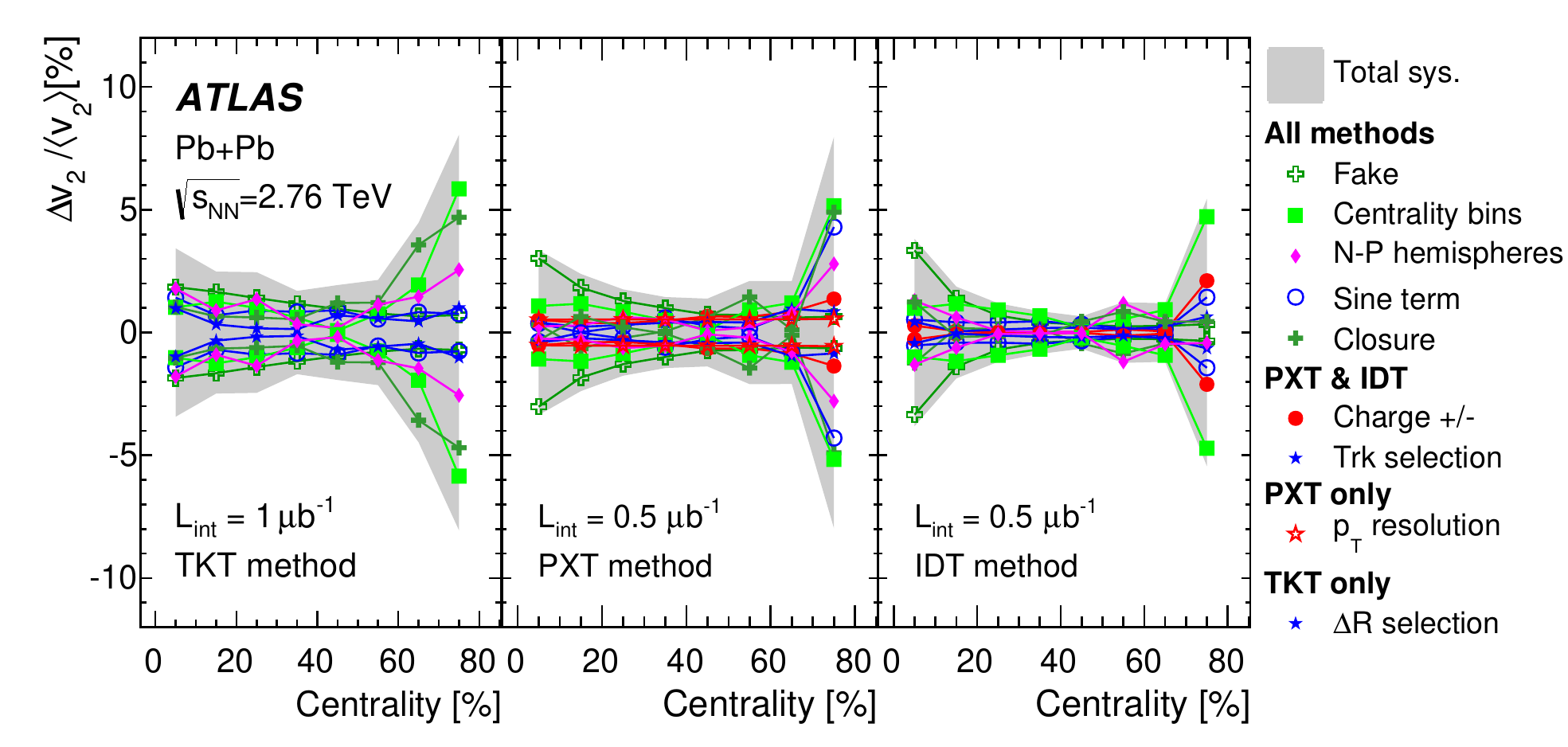}
\caption[systematics]{\label{fig:system} Contributions to the
    relative systematic uncertainty on the elliptic flow, $\Delta \vv
    / \langle \vv\rangle$, as a function of centrality for $|\eta|<1$
    with the TKT (left), PXT (centre) and IDT (right)
    methods. The integration limits for the three methods are 0.07, 0.1, 0.5\,\GeV, respectively.
    The total
    uncertainty is indicated by the shaded area. The individual
    contributions, are described in the legend and explained in the
    text.}
\end{center}
\end{figure*}

The different contributions to the total systematic uncertainty on the
integrated $\vv$ for  $|\eta|<1$ are shown in Fig.~\ref{fig:system} 
and  summarized in Table~\ref{tab:system}  for the three tracking methods.
The total systematic uncertainties are determined by adding in
quadrature all the individual contributions and are treated as
$\pm 1\sigma$ uncertainties.

\renewcommand{\tabcolsep}{1mm}
\begin{table}[!t]
  \centering
  \vspace*{2.8mm}
  \begin{tabular}{|c|ccccc|}
    \hline
     & \multicolumn{5}{|c|}{Centrality bin } \\
    \cline{2-6}
    Source &0--10\%&10--20\%&20--60\%&60--70\%&70--80\%\\
    \cline{2-6}
           &  \multicolumn{5}{|c|}{TKT $\pT > 0.07$\,\GeV} \\
    \hline
    MC Statistics  & 0.1 & 0.1 & <0.2 & 0.3 & 1 \\
    Fake tracks & 2    & 2    & 1--2   & 1    & 1 \\
    Centrality bins  & 1    & 1.5 & <1    & 2    & 6 \\
    N-P $\eta$ regions  & 2    & 1    & <1.5 & 1    & 2.5 \\
    Sine term   & 1.5 & 1  & 1   & 1  & 1 \\
    Closure      & 1.5 & 1    & <2   & 3.5     & 5 \\
    $\Delta R$ & 1 & 0.5  & <1   & 0.5  & 1 \\
    Total          & 3.5 & 3.2  & <3.2  & 4  & 8 \\
    \hline
    &  \multicolumn{5}{|c|}{PXT $\pT > 0.1$\,\GeV} \\
    \hline
    MC Statistics         & 0.1 & 0.1 & <0.2 & 0.3 & 1 \\
    Fake tracks      & 3    & 2    & <1.5 & 0.5 & 0.5 \\ 
    Centrality bins  & 1   & 1.5 & <1 & 1.5  & 5 \\
    N-P $\eta$ regions & 0.5 & 0.5 & < 0.5 & 1  & 3 \\
    Sine term          & 0.5 & 0 & <0.5 & 1 & 4 \\
    Closure             & 1      &   1 & $<$2 & 0 & 5 \\
    Charge $\pm$  & 0.5 & 0.5 & <1 & 1 & 1.5 \\
    Track selection & 0.5 & 0.5 & <0.5 & 1 & 1 \\
    \pT\ resolution & 0.5 & 0.5 & 0.5 & 0.5 & 0.5 \\
    Total                 & 3    &  2  &  <2  & 2  & 8 \\
    \hline
    &  \multicolumn{5}{|c|}{IDT $\pT > 0.5$\,\GeV} \\
    \hline
    MC Statistics          & 0.1  & 0.1 & <0.2 & 0.3   & 1 \\
    Fake tracks       & 3.5  &  1.5 & <1   & 0.2 & 0.2\\ 
    Centrality bins   & 1     & 1.5  & <1   & 1      & 5 \\
    N-P $\eta$ regions  & 1.2 & 1 &<1.5 & 0.5  & 0.5 \\
    Sine term           & 0.5   & 0.5  & 0.5  & 0.5  & 1.5 \\
    Closure              & 1.5   & 0.5  & <1  & 0.5  & 0.5 \\
    Charge $\pm$   & 0.2   & 0.2  & 0.2  & 0.2 & 2.2 \\
    Track selection  & 0.5   & 0     & <0.5 & 0.2 & 1 \\
    Total                  &  3.5  & 2    & <1.5   &  1 & 5.5 \\
    \hline
  \end{tabular}  
  \caption{Summary of the systematic uncertainties as percentages of the integrated $\vv$ value for charged particles with $|\eta|<1$ and different collision centrality bins. }
  \label{tab:system}
\end{table}

 \begin{figure*}[!b]
\begin{center}
\includegraphics[width=0.9\textwidth]{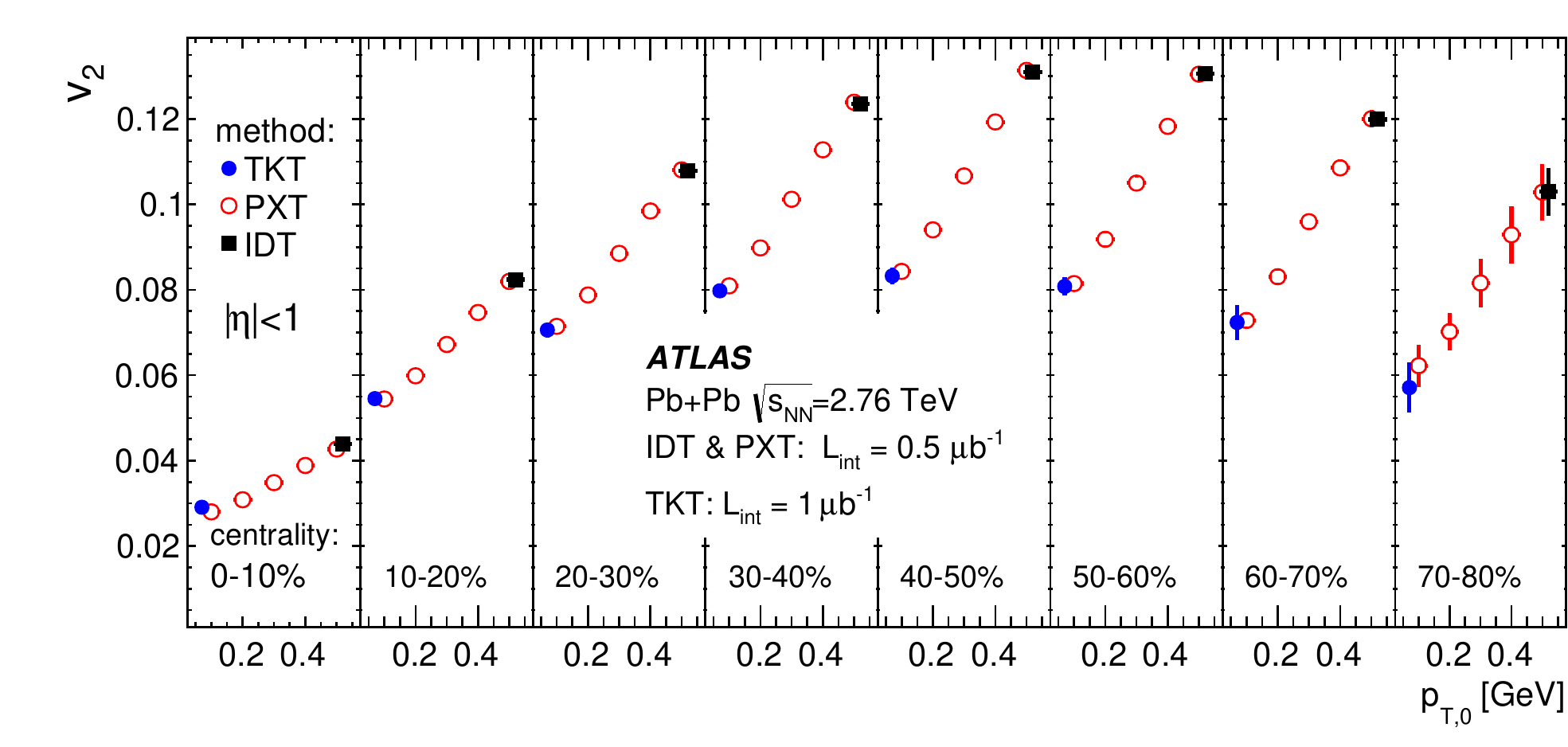}
\caption[consistency]{\label{fig:consist} Elliptic flow $\vv$
  integrated over transverse momentum $\pT>\pTzero$ as a function of 
  $\pTzero$ for different centrality intervals, obtained with different
  charged-particle reconstruction methods: the tracklet (TKT) method with
  $\pTzero = 0.07$\,\GeV, the pixel track (PXT) method with 
  $\pTzero \ge 0.1$\,\GeV\ and the ID track (IDT) method with 
  $\pTzero = 0.5$\,\GeV\ as described in the legend. Error bars show
  statistical and systematic uncertainties added in quadrature. }
\end{center}
\end{figure*}

\section{Results}
\label{sec:results}
Figure~\ref{fig:consist} shows the centrality dependence of $\vv$
integrated over $|\eta|<1$. 
For the TKT method, $\vv$ is integrated over $\pT>0.07$\,\GeV.
For the PXT method, $\vv$ is integrated over $\pTzero < \pT < 5$\,\GeV\ 
and \pTzero\ is varied from 0.1\,\GeV\ to 0.5\,\GeV\ in steps of 0.1\,\GeV. 
Also shown is the $\vv$ value obtained from the IDT method integrated
over $0.5 < \pT < 5$\,\GeV. 
The TKT method with $\pTzero = 0.07$\,\GeV\ gives results consistent
with the $\vv$ values obtained with the PXT method with 
$\pTzero =0.1$\,\GeV, as could be expected due to the very low
charged-particle density and small $\vv$ signal in the 
momentum range below 0.1\,\GeV. 
This indicates that there is no need to extrapolate the measurements
obtained with tracklets down to $\pT = 0$ in order to obtain a
reliable estimate of $\vv$ integrated over the whole kinematic range
in \pT. 
Furthermore, for the PXT method such an extrapolation would result in
a very small correction to the measured integrated flow, well within
the uncertainties of the measurement. 
This is in contrast to the integrated $\vv$ with \pTzero\ chosen at
higher values, as also shown in Fig.~\ref{fig:consist}.
It can be seen that the integrated $\vv$ increases almost linearly
with \pTzero\ for $\pTzero > 0.1$\,\GeV. 
Good agreement between the PXT and IDT methods is observed at 
$\pTzero= 0.5$\,\GeV.

\begin{figure}[!h]
\begin{center}
\includegraphics[width=\colfig]{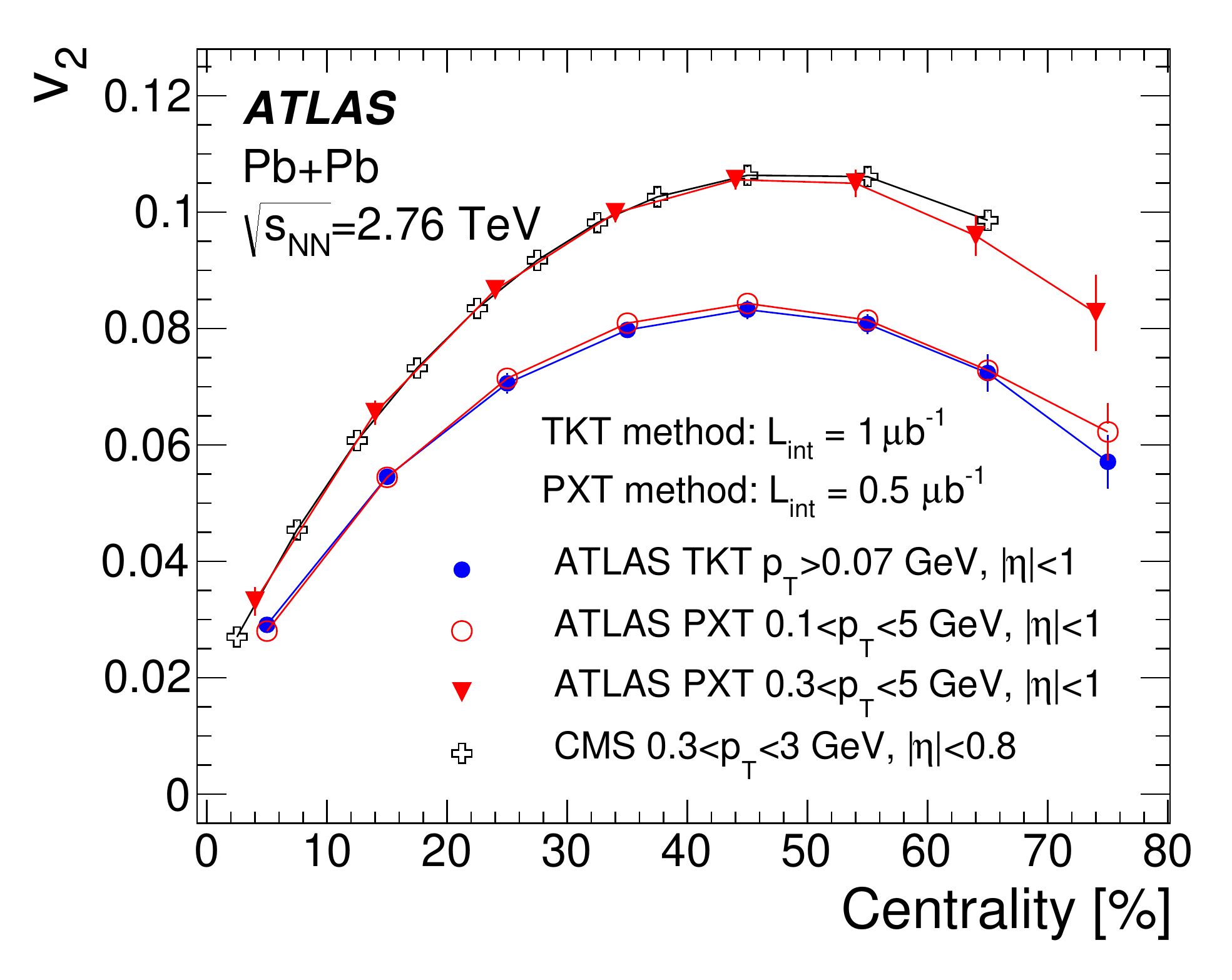}
\caption[centrality]{\label{fig:centrdep} Centrality dependence of
 elliptic flow,  $\vv$, measured for $|\eta|<1$ and integrated over transverse
  momenta, \pT, for different charged-particle reconstruction methods as described in the legend.
  Also shown are $\vv$ measurements by CMS integrated over
  $0.3<\pT<5$\,\GeV\ and $|\eta|<0.8$ \cite{Chatrchyan:2012ta} (open crosses). Error bars show
  statistical and systematic uncertainties added in quadrature.  }
\end{center}
\end{figure}

In Fig.~\ref{fig:centrdep}, the results of this analysis are compared to the integrated $\vv$
measured by 
CMS \cite{Chatrchyan:2012ta} with $\pTzero = 0.3$\,\GeV.
In this comparison, the sensitivity to \pTzero\ is clearly visible. 
A systematically larger $\vv$ is observed for the higher value of \pTzero\ 
as a consequence of the strong increase of $\vv$ with increasing \pT.
\begin{figure*}[!t]
\begin{center}
  \vspace*{2.8mm}
\includegraphics[width=0.9\textwidth]{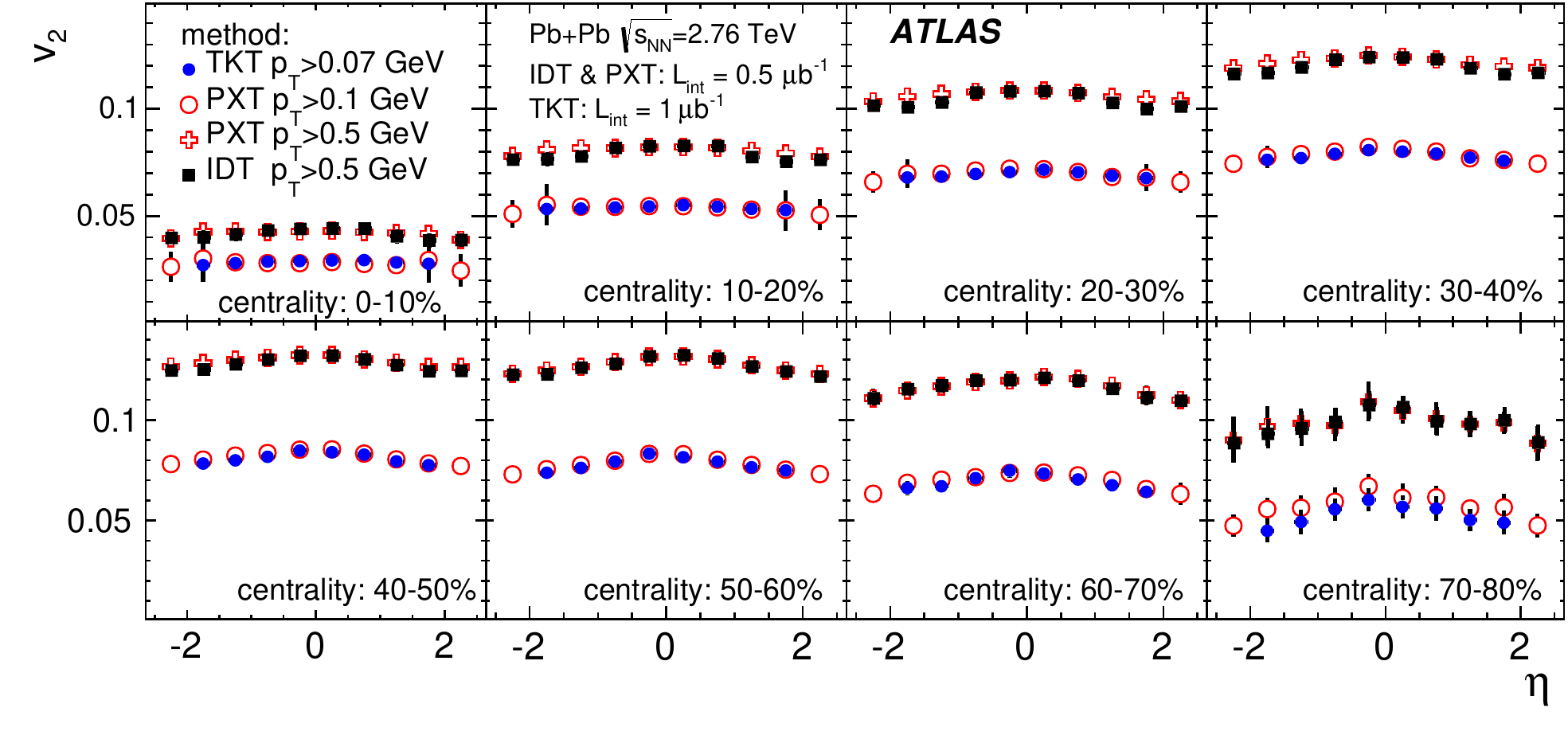}
\caption[pseudorapidity]{\label{fig:etadep} Pseudorapidity, $\eta$,
  dependence of elliptic flow, $\vv$, integrated over transverse
  momentum, $\pT$, for different charged particle reconstruction
  methods and different low-$\pT$ thresholds in different centrality
  intervals as indicated in the legend. Error bars show statistical and systematic uncertainties
added in quadrature.}
\end{center}
\end{figure*} 
 
\begin{figure}[t]
  \begin{center}
    \includegraphics[width=\colfig]{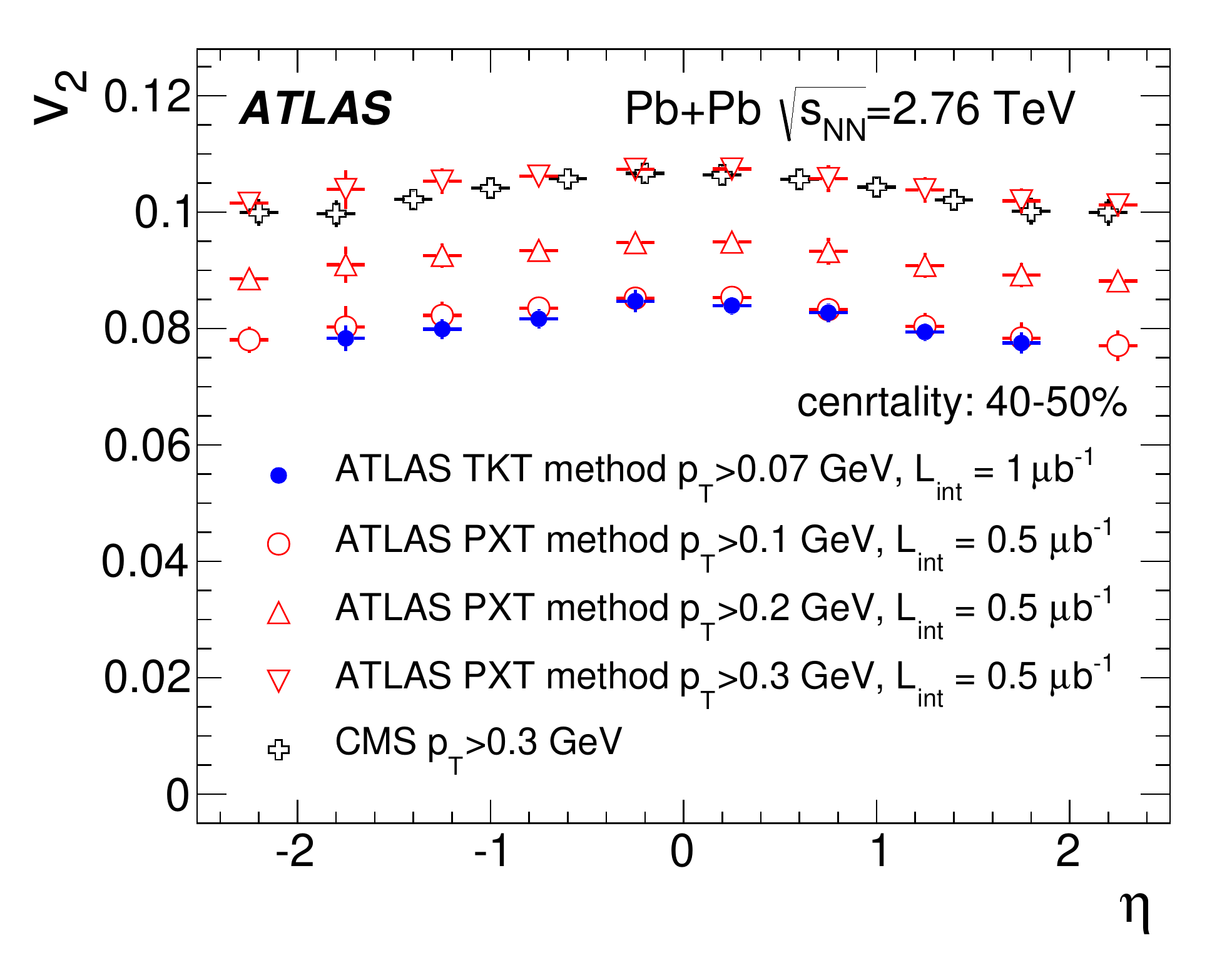}
    \caption[pseudorapidity comparison]{\label{fig:expcomp}
      Comparison of the pseudorapidity, $\eta$, dependence of elliptic
      flow, $\vv$, integrated over transverse momentum, \pT, for different
      low-\pT\ thresholds, as indicated in the legend,  in the 40--50\%
      centrality interval from the ATLAS and CMS experiments. 
      Error bars show statistical and systematic uncertainties added in quadrature. }
  \end{center}
\end{figure}

The \eta\ dependence of the \pT-integrated $\vv$ provides useful
constraints on the initial conditions of heavy-ion collisions used in
model descriptions of the system's evolution (see, e.g.,
Refs~\cite{Romatschke:2009im,Hirano:2005xf}).
Figure~\ref{fig:etadep} shows the $\eta$ dependence of the
\pT-integrated $\vv$. 
As already shown in Fig.~\ref{fig:centrdep}, the difference between
the results obtained with \pTzero\ values of 0.07\,\GeV\ and
0.1\,\GeV\ is very small and the two measurements agree within
uncertainties. 
The results obtained using the PXT and IDT methods for the same
\pTzero\ are also consistent. 
The \eta\ dependence of the integrated $\vv$ is very weak. 
A decrease with increasing $|\eta|$ of about 5--10\% is seen.
A comparison with the results from the CMS
experiment~\cite{Chatrchyan:2012ta} is shown in Fig.~\ref{fig:expcomp}
for the 40--50\% centrality interval. 
The ATLAS measurements performed with the PXT method were integrated
over \pT\ for different \pTzero\ values, including one adjusted to
match that used by CMS. 
The results agree, within uncertainties, provided the same \pTzero\ is used.  

The different upper limits in the \pT\ integration, 3\,\GeV\ for CMS and
5\,\GeV\ for ATLAS, have negligible effect on the integrated $\vv$ value.  
A systematic decrease in $\vv$ with decreasing \pTzero\ is observed as
expected from the linear dependence of \vv\ on \pT\ for $\pT \approx 0$.
The decreasing value of \pTzero\ together with that of $\vv$ makes the
integration over the full \pT\ range less sensitive to the
uncertainties in the extrapolation down to $\pT = 0$.
\begin{figure*}[ht!]
\begin{center}
\includegraphics[width=0.9\textwidth]{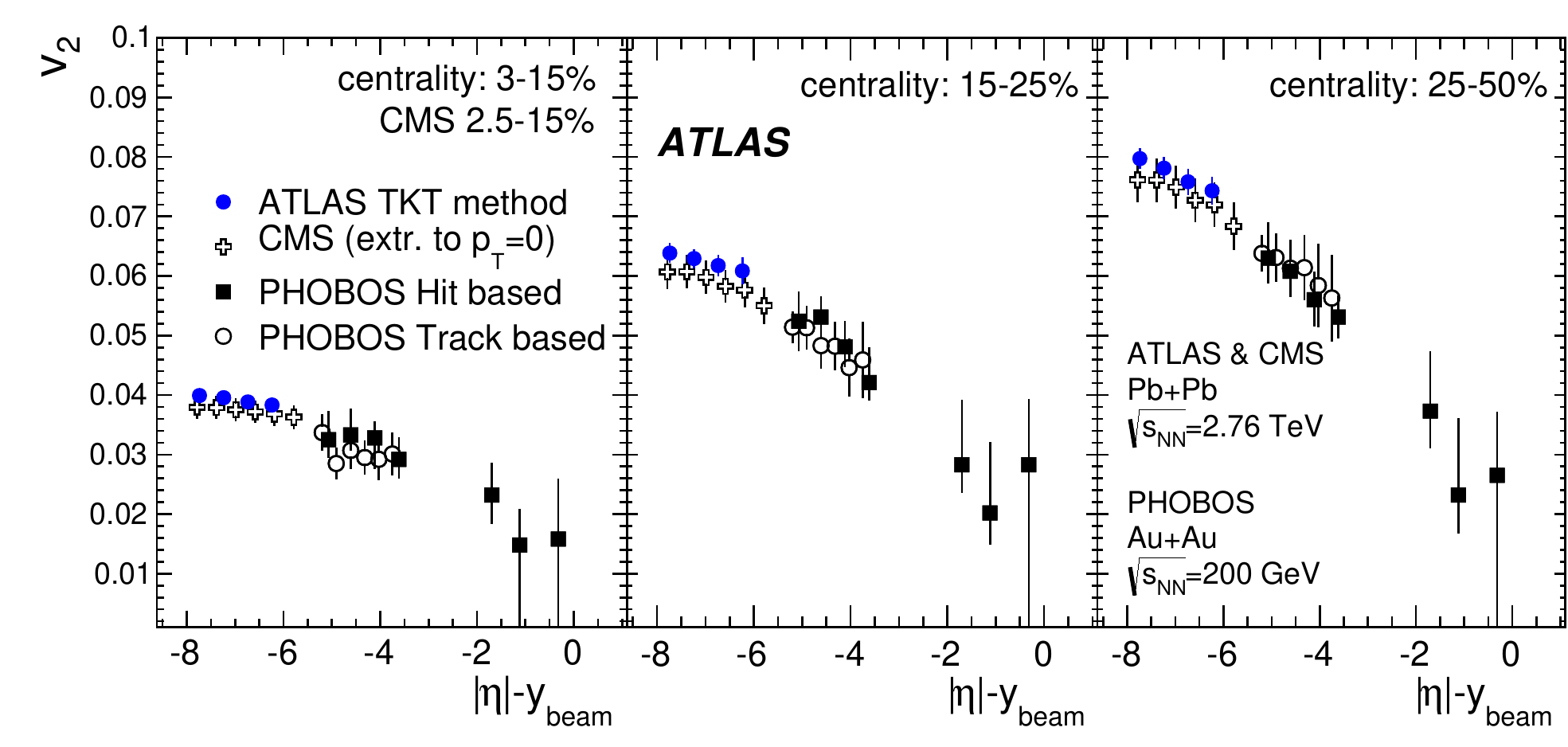}
\caption[limiting frag]{\label{fig:limit} Integrated elliptic flow, $\vv$, as a
  function of $|\eta|-y_{\mathrm{beam}}$ for three centrality intervals 
indicated in the legend, measured by the ATLAS and CMS experiments for Pb+Pb
collisions at 2.76\,\TeV\ and by the PHOBOS experiment for Au+Au collisions at 200\,\GeV. 
The CMS result is obtained by averaging the $\vv(\pT)$ with the charged particle spectra
over the range $0 < \pT < 3$\,\GeV.
Error bars show statistical and systematic uncertainties
added in quadrature.}
\end{center}
\end{figure*}

The large acceptance in \eta\ of the ATLAS and CMS experiments makes
it possible to study whether the observation of the extended
longitudinal scaling of $\vv$, when viewed in the rest frame of one of
the colliding nuclei, reported by the PHOBOS experiment at
RHIC~\cite{Back:2004je,Back:2004mh}, holds at the much higher LHC
energy. 
The PHOBOS measurements of elliptic flow over a range of Au+Au
collision energies, $\sqrt{s_{\mathrm{NN}}} = 19.6$, 62.4, 130 and
200\,\GeV, showed energy independence of the integrated $\vv$ as a
function of $|\eta|-y_{\mathrm{beam}}$, where  
$y_{\mathrm{beam}}=\ln{(\sqrt{s_{\mathrm{NN}}}/m)}$ is the beam
rapidity and $m$ is the proton mass. 
A similar effect was also observed in charged-particle
densities~\cite{Back:2004je} and is known as limiting
fragmentation~\cite{Benecke:1969sh}. 
In Fig.~\ref{fig:limit}, the integrated $\vv$ is plotted as a function of
$|\eta|-y_{\mathrm{beam}}$ and compared to the PHOBOS results for
three centrality bins matching those used by PHOBOS. 
The PHOBOS results are obtained with the event-plane method for
charged particles with a low-\pT\ limit of 0.035\,\GeV\ at
mid-rapidity and of 0.004\,\GeV\ around the beam
rapidity~\cite{Back:2005hs}. 
The CMS data \cite{Chatrchyan:2012ta} obtained with the event-plane
method are also shown. 
The CMS measurement represents $\vv$ integrated over \pT\ from 0 to 
3\,\GeV. 
This measurement was obtained by extrapolating $\vv(\pT)$ measured
for $\pT > 0.3$\,\GeV\ and the charged-particle spectra down to
$\pT=0$ under the assumption that $\vv(\pT=0)=0$ and with the
charged-particle yield constrained by the measured
${\mathrm{d}}\Nch/{\mathrm{d}}\eta$ distribution~\cite{Chatrchyan:2011pb}.

The ATLAS and CMS results are shifted systematically by up to 5\% while the uncertainties of ATLAS measurement are of the order of 3-5\% and CMS are or the order of 5-6\%.
However, as can be seen from the figure, there is no overlap in
$|\eta|-y_{\mathrm{beam}}$ between the PHOBOS and LHC data, so a
direct comparison with the low-energy data is not possible. 
Nevertheless, it can be concluded, keeping in mind the relatively
large uncertainties in the low-energy results, that the extrapolation
of the trend observed at RHIC to the LHC energy appears to be
consistent with the LHC measurements, although the dependence on 
$|\eta|-y_{\mathrm{beam}}$ may be weaker at the LHC energy.

\section{Summary and conclusions}
Measurements of the integrated elliptic flow of charged particles in
Pb+Pb collisions at $\sqrt{s_{\mathrm{NN}}} = 2.76$\,\TeV\ are
presented by the ATLAS experiment at the LHC. 
The elliptic anisotropy parameter $\vv$ is measured with the
event-plane method over a broad range of collision centralities
(0--80\%). 
The kinematic range in pseudorapidity extends out to $|\eta|=2.5$, and
in \pT\ down to 0.07\,\GeV. 
This low-\pT\ region is reached by using a tracklet reconstruction
algorithm to analyze about $1\,\mu \mathrm{b}^{-1}$ of data taken with
the solenoid field turned off. 
Other track reconstruction methods with low-\pT\ thresholds of
0.1\,\GeV\ and 0.5\,\GeV\ respectively, are exploited in order to
verify the tracklet measurement and provide results that can be
directly compared to other LHC measurements. 
The values of $\vv$ integrated from $\pT = 0.07$\,\GeV\ and from
$\pT=0.1$\,\GeV\ agree with each other to better than 4\%, 
which proves that the contribution from  tracks with \pT\ below 0.1 \GeV\ is negligible and therefore no model-dependent extrapolation to \pT=0 is necessary.

The \pT-integrated elliptic flow as a function of collision centrality
shows a clear dependence on \pTzero, both within the present
measurements and in comparison to the CMS results obtained with higher 
low-\pT\ limits. 
The integrated elliptic flow increases with centrality, reaching a
maximum of 0.08 for mid-central collisions (40--50\%) and then
decreases to about 0.02 for the most central collisions.

The pseudorapidity dependence of the \pT-integrated $\vv$ is very
weak, with a slight decrease in $\vv$ as $|\eta|$ increases. 
The results are in agreement with the CMS measurements covering the
same $\eta$ range, provided the same low-\pT\ cutoff is used. 
The integrated $\vv$ transformed to the rest frame of one of the
colliding nuclei is compared to the lower-energy RHIC data. 
Although a direct comparison is not possible due to the
non-overlapping kinematic regions, the general trend observed in the
RHIC energy regime seems consistent with the LHC measurements, while
the latter may have a weaker dependence on pseudorapidity.

\section{Acknowledgements}

We thank CERN for the very successful operation of the LHC, as well as the
support staff from our institutions without whom ATLAS could not be
operated efficiently.

We acknowledge the support of ANPCyT, Argentina; YerPhI, Armenia; ARC,
Australia; BMWF and FWF, Austria; ANAS, Azerbaijan; SSTC, Belarus; CNPq and FAPESP,
Brazil; NSERC, NRC and CFI, Canada; CERN; CONICYT, Chile; CAS, MOST and NSFC,
China; COLCIENCIAS, Colombia; MSMT CR, MPO CR and VSC CR, Czech Republic;
DNRF, DNSRC and Lundbeck Foundation, Denmark; EPLANET, ERC and NSRF, European Union;
IN2P3-CNRS, CEA-DSM/IRFU, France; GNSF, Georgia; BMBF, DFG, HGF, MPG and AvH
Foundation, Germany; GSRT and NSRF, Greece; ISF, MINERVA, GIF, I-CORE and Benoziyo Center,
Israel; INFN, Italy; MEXT and JSPS, Japan; CNRST, Morocco; FOM and NWO,
Netherlands; BRF and RCN, Norway; MNiSW and NCN, Poland; GRICES and FCT, Portugal; MNE/IFA, Romania; MES of Russia and ROSATOM, Russian Federation; JINR; MSTD,
Serbia; MSSR, Slovakia; ARRS and MIZ\v{S}, Slovenia; DST/NRF, South Africa;
MINECO, Spain; SRC and Wallenberg Foundation, Sweden; SER, SNSF and Cantons of
Bern and Geneva, Switzerland; NSC, Taiwan; TAEK, Turkey; STFC, the Royal
Society and Leverhulme Trust, United Kingdom; DOE and NSF, United States of
America.

The crucial computing support from all WLCG partners is acknowledged
gratefully, in particular from CERN and the ATLAS Tier-1 facilities at
TRIUMF (Canada), NDGF (Denmark, Norway, Sweden), CC-IN2P3 (France),
KIT/GridKA (Germany), INFN-CNAF (Italy), NL-T1 (Netherlands), PIC (Spain),
ASGC (Taiwan), RAL (UK) and BNL (USA) and in the Tier-2 facilities
worldwide.

\bibliographystyle{modspphys}
\bibliography{Paper}

\section*{Appendix}
In the low-\pT\ region, the track reconstruction efficiency depends
strongly on the particle type.
This information is important for comparison of measurements with
theory predictions in which the elliptic flow depends on the particle
type.

\begin{figure}[!htb]
  \begin{center}
    \includegraphics[width=\colfig]{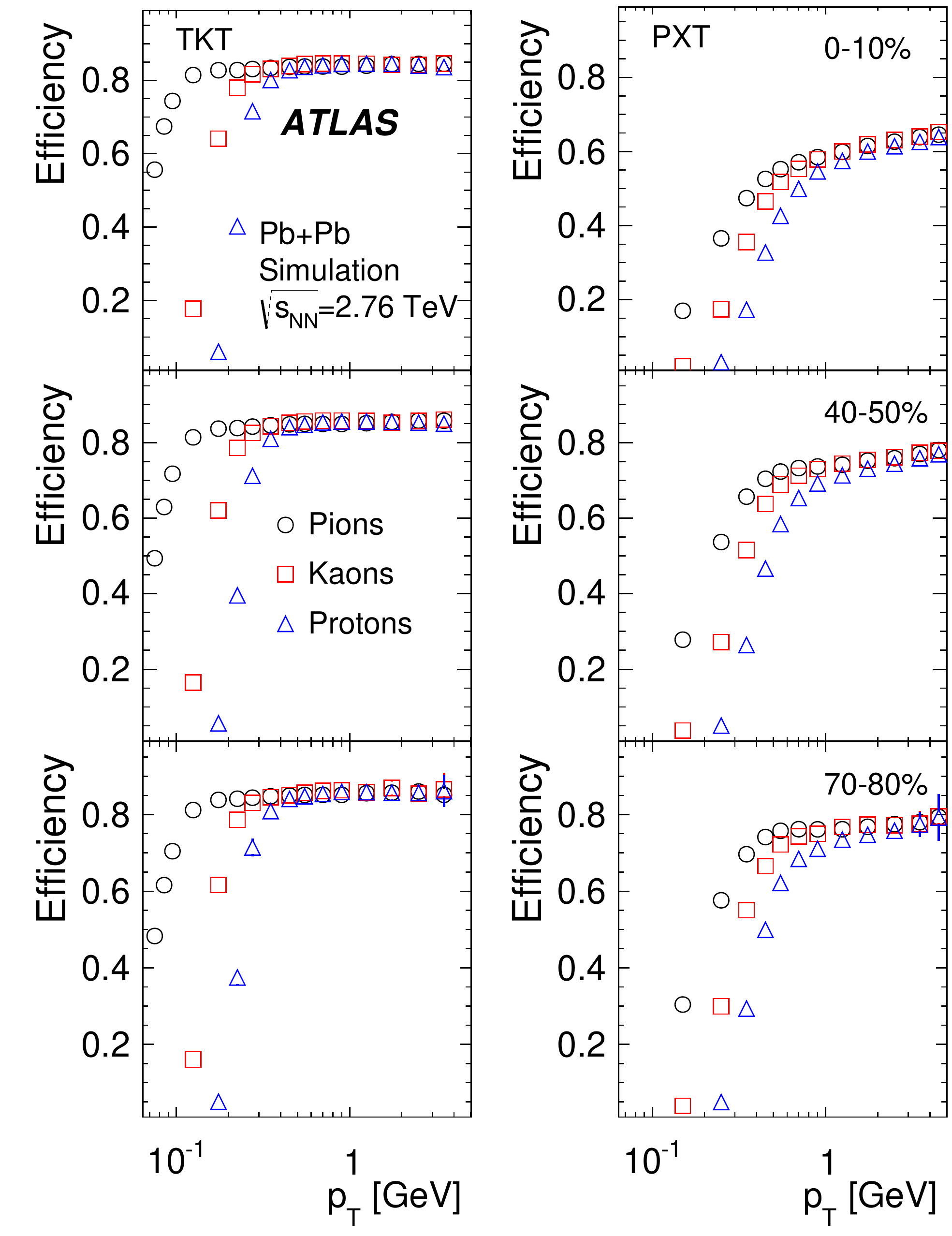}
    \caption[pixel performance]{
      The transverse momentum, \pT, dependence of the TKT 
      (left) and PXT (right) track reconstruction efficiency for $\pi^\pm$,
      $K^\pm$ and $p^\pm$ in the pseudorapidity range $|\eta|<1$ 
      for three centrality intervals, as indicated in the legend.}
    \label{fig:eff_id} 
  \end{center}
\end{figure}

The efficiency of the PXT and TKT methods in reconstructing tracks
with $|\eta|<1$ generated as $\pi^\pm$, $K^\pm$, $p$, and $\bar{p}$ in
MC simulation is shown in Fig.~\ref{fig:eff_id} as a function of \pT.  
Large differences in efficiency are observed for the PXT method at
\pT\ below about 1\,\GeV\ and for the TKT method at \pT\ below about 
0.4\,\GeV. 
Above these values, the reconstruction efficiency is independent of
particle type.  The efficiency is lowest for $p$ and $\bar{p}$. 
For the TKT method, which is most relevant at low \pT, the
efficiency for reconstructing protons drops to zero below 0.2\,\GeV.

\flushend

\clearpage
\begin{flushleft}
{\Large The ATLAS Collaboration}

\bigskip

G.~Aad$^{\rm 84}$,
B.~Abbott$^{\rm 112}$,
J.~Abdallah$^{\rm 152}$,
S.~Abdel~Khalek$^{\rm 116}$,
O.~Abdinov$^{\rm 11}$,
R.~Aben$^{\rm 106}$,
B.~Abi$^{\rm 113}$,
M.~Abolins$^{\rm 89}$,
O.S.~AbouZeid$^{\rm 159}$,
H.~Abramowicz$^{\rm 154}$
% ...
\bigskip
\\
% ...
$^{11}$ Institute of Physics, Azerbaijan Academy of Sciences, Baku, Azerbaijan\\
% ...
$^{84}$ CPPM, Aix-Marseille Universit{\'e} and CNRS/IN2P3, Marseille, France\\
% ...
$^{89}$ Department of Physics and Astronomy, Michigan State University, East Lansing MI, United States of America\\
% ...
$^{106}$ Nikhef National Institute for Subatomic Physics and University of Amsterdam, Amsterdam, Netherlands\\
% ...
$^{112}$ Homer L. Dodge Department of Physics and Astronomy, University of Oklahoma, Norman OK, United States of America\\
$^{113}$ Department of Physics, Oklahoma State University, Stillwater OK, United States of America\\
% ...
$^{116}$ LAL, Universit{\'e} Paris-Sud and CNRS/IN2P3, Orsay, France\\
% ...
$^{152}$ Institute of Physics, Academia Sinica, Taipei, Taiwan\\
% ...
$^{154}$ Raymond and Beverly Sackler School of Physics and Astronomy, Tel Aviv University, Tel Aviv, Israel\\
% ...
$^{159}$ Department of Physics, University of Toronto, Toronto ON, Canada\\
% ...
\end{flushleft}


\end{document}